\documentclass[twocolumn,astrosymb]{aastex7}

\newcommand{\Msun}{$\mathrm{M_\odot}$}
\newcommand{\Mmsto}{$\rm M_{MSTO}$}
\newcommand{\Mrel}{$m_{\rm{rel}}$}
\newcommand{\age}{$\tau_{\rm{rel}}$}
\usepackage[T1]{fontenc}

\begin{document}

\title{The Distribution of Blue Straggler Stars in the Color-Magnitude Diagrams of Old Open Clusters}
\correspondingauthor{Evan Linck}
\author[0009-0006-7474-7889]{Evan Linck}
\affiliation{Department of Astronomy, University of Wisconsin-Madison, 475 N. Charter St., Madison, WI 53706, USA}
\email{elinck@wisc.edu}
\author[0000-0002-7130-2757]{Robert D. Mathieu}
\affiliation{Department of Astronomy, University of Wisconsin-Madison, 475 N. Charter St., Madison, WI 53706, USA}
\email{mathieu@astro.wisc.edu}

\received{2026 February 22}

\begin{abstract}

We examine the blue straggler star (BSS) populations of six old ($\geq$4 Gyr) open clusters: M67, NGC 188, NGC 6791, Berkeley 32, Berkeley 39, and Trumpler 19. We find that 50\% of BSSs have color-magnitude diagram (CMD) locations corresponding to single stars in the final third of their main-sequence lifetimes. This build-up of BSSs near the terminal-age main sequence (TAMS) is primarily, but not solely, driven by more massive BSSs. Eleven of the BSSs have white dwarf companions with measured cooling ages; their evolution age distributions indicate that more massive BSSs typically form far from the zero-age main sequence, whereas lower mass BSSs can form at every evolutionary age. We show that inferred core helium amounts (above primordial) of late-evolution-age BSSs correspond to the core helium fused by cluster main-sequence stars near the turnoffs. We also find that the masses of asymptotic giant branch (AGB) mass-transfer BSSs require evolved main-sequence accretors and conservative mass transfer. These findings indicate that helium enrichment of progenitor accretors leads to the prevalence of BSSs near the TAMS. We further classify the evolutionary stages of the progenitor donors in M67 and NGC 188 and find mass transfer during the AGB accounts for at least half of the BSSs. We trace how the main-sequence binary population of NGC 188 evolves, and find that only 30--40\% of interacting binaries create BSSs and that progenitor orbits must change to match current BSS periods. 

\end{abstract}
 
\section{Introduction}\label{sec:introduction}

Blue straggler stars (BSSs) are main-sequence stars that have gained mass through interactions with companion stars. They are important tracers of initial evolutionary pathways of binary stars \citep{mathieuBlueStragglersFriends2025}.

Binary stars are thought to be the progenitors of BSSs through three formation mechanisms: mass transfer between companion stars \citep{mccreaExtendedMainSequenceStellar1964,boffinRochelobeFillingFactor2014}, mergers of binary stars from angular momentum loss \citep{andronovMergersClosePrimordial2006,peretsTRIPLEORIGINBLUE2009}, and collisions of stars during resonant binary encounters \citep{leonardStellarCollisionsGlobular1989,leighAnalyticTechniqueConstraining2011}, with mass transfer likely responsible for up to 75\% of BSS formation \citep{gosnellIMPLICATIONSFORMATIONBLUE2015,leinerCensusBlueStragglers2021}. 

In cluster environments, classical BSSs stand out from other cluster members on the color-magnitude diagram (CMD) as stars that are brighter and/or bluer than the cluster main sequence. Specifically, in this paper we consider only BSSs between the zero-age main sequence (ZAMS) and terminal-age main sequence
(TAMS) whose locations distinguish them from single stars or photometric binaries in the associated cluster population. Thus, for example, we do not include blue lurkers---binary interaction products embedded within the cluster main sequence \citep{leinerBlueLurkersHidden2019}. With the advent of cluster membership lists derived from the high-precision proper-motion, parallax, and photometry measurements of the \textit{Gaia} space observatory, numerous studies have analyzed the CMD locations of such BSSs, with particular emphasis on their inferred masses \citep[e.g,.][]{jadhavBlueStragglerStars2021,  leinerCensusBlueStragglers2021, rainBinaryOriginBlue2024} and ages \citep[e.g,.][]{carrasco-varelaMainSequenceBinary2025}. 

A consequence of gaining mass is a variety of relevant ages for BSSs. All have the same \textit{formation age} as the cluster stars, since their progenitor binaries formed with the cluster. Two additional time scales exist for each BSS: the \textit{transformation age}---the time since the completion of formation of the BSS---and the \textit{evolution age}---the age of a single star with the same mass and interior structure \citep{sunWOCS5379Detailed2021, mathieuBlueStragglersFriends2025}. 

Such ages, combined with masses and orbital properties, are crucial for deducing the formation pathways of BSSs and physical details of the initial evolution of low-mass binary stars---such as progenitor masses, mass ratios, and orbital characteristics, mass-transfer efficiencies, and stellar structures post-interaction. This work examines the ages, masses, and orbits of BSSs in old ($\geq 4$ Gyr) open clusters. BSSs in open clusters are excellent case studies of formation and binary evolution as 1) each cluster BSS population is at the same distance; 2) tight constraints can be placed on formation ages; 3) proximity and lower density environments have enabled long-time-baseline radial-velocity monitoring that provide orbit solutions; 4) similarly, photometric monitoring yields rotation periods; 5) ultraviolet (UV) surveys define white dwarf (WD) companions; and 6) abundances are well determined at or near solar metallicity. Physically, the progenitor binaries have masses that straddle major regions of stellar structure changes, such as the shifts from the proton-proton chain to the carbon-nitrogen-oxygen cycle dominating core hydrogen fusion and from convective to radiative envelopes. Finally, the cores of the subgiant and red giant branch (RGB) stars are degenerate, which causes the cores to have very similar masses at the same state of evolution regardless of exact stellar mass.

As a canonical example to motivate this paper, the BSSs of the open cluster NGC 188 (6.6 Gyr, CMD is shown in Figure~\ref{fig:cmds}) are quite well studied \citep[e.g.,][]{mathieuBinaryStarFraction2009,geller_mass_2011,gosnellIMPLICATIONSFORMATIONBLUE2015,subramaniamHotCompanionBlue2016,gosnellConstrainingMasstransferHistories2019,narayanWIYNOpenCluster2026}. Several key properties stand out. First, only one of the most luminous BSSs is found on the ZAMS; in fact most of the more luminous BSSs are near the TAMS. Second, some lower luminosity BSSs do lie on or near the ZAMS. Third, BSSs are present at luminosities below the main-sequence turnoff (MSTO)---defined in this work as the end of core hydrogen burning---and blueward of the main sequence. The goal of this paper is to analyze and interpret these distributions in a set of old open clusters.

Specifically, this work examines the distribution of masses and transformation and evolutionary ages of BSSs in 6 open clusters older than 4 Gyr: M67, Trumpler 19, Berkeley 32, Berkeley 39, NGC 188, and NGC 6791 with the goal of extending our understanding of BSS formation and initial low-mass binary evolution. We especially use the BSSs of M67 and NGC 188 for detailed studies given the large bodies of existing research on these two clusters. In Section \ref{sec:BSS}, we define our BSS samples and describe our methodology for determining BSS characteristics (e.g., mass and ages). In Section \ref{sec:results}, we present our observational and modeling results, including trends found among BSS populations. In Section \ref{sec:discussion}, we discuss possible causes of these trends and their implications for our understanding of binary star evolution. In Section~\ref{sec:summary}, we summarize our findings and conclusions.

\section{Blue Straggler Stars in Old Open Clusters}\label{sec:BSS}

\subsection{Cluster Selection and Membership Lists}
\begin{figure*}
    \centering
    \includegraphics[width=\linewidth]{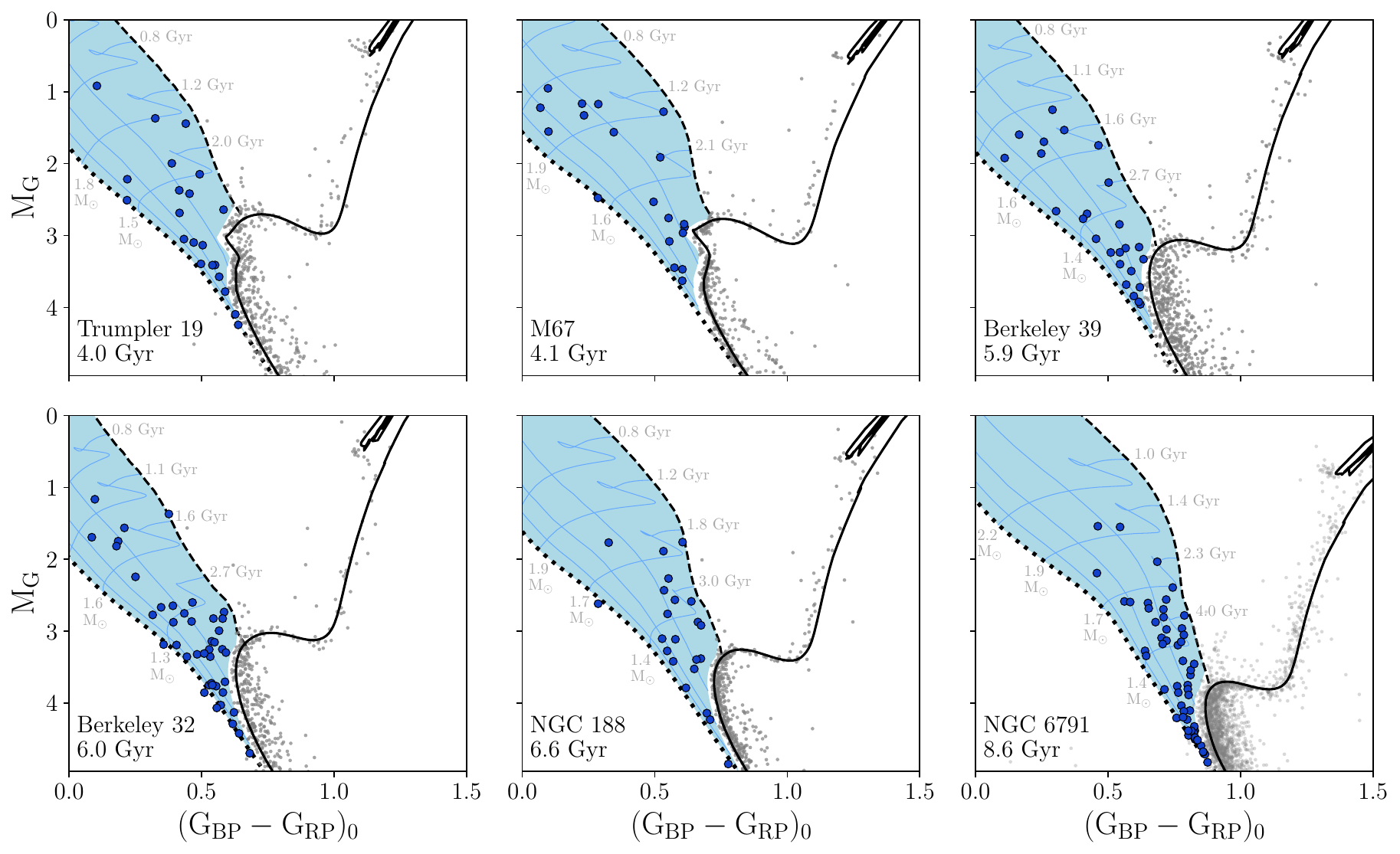}
    \caption{The extinction- and distance-corrected CMDs of the six open clusters in this study: Trumpler 19 (4.0 Gyr), M67 (4.1 Gyr), Berkeley 39 (5.9 Gyr), Berkeley 32 (6.0 Gyr), NGC 188 (6.6 Gyr), and NGC 6791 (8.6 Gyr). Each cluster is shown with its best-fit MIST isochrone (solid line) from Section \ref{sec:isochrone} along with the corresponding ZAMS (dotted line) and TAMS (dashed line). The BSS region of each cluster is bounded by the ZAMS, TAMS, and a clear gap of at least 0.02 mag in color from any group of stars on the main sequence. BSSs stars are plotted as blue circles whereas other cluster members are plotted as grey circles. For visual reference, the BSS region is divided in regions by light-blue lines between the ZAMS and TAMS that denote the position of single stars on the CMD that are 25\%, 50\% and 75\% of their main-sequence lifetime (with the ZAMS and TAMS representing 0\% and 100\%, respectively). Additionally, evolutionary tracks of stars that are 1.25, 1.5, 1.75, and 2 times the \Mmsto\;are plotted in light-blue along with their masses and total main-sequence lifetimes.}
    \label{fig:cmds}
\end{figure*}

We examined the CMD of every open cluster that was reported to be at least 4 Gyr old by at least one of \citet{diasUpdatedParameters17432021}, \citet{huntImprovingOpenCluster2023}, or \citet{cantat-gaudinPaintingPortraitGalactic2020}:  Berkeley 17, 18, 20, 32, 36, and 78; NGC 188, 1193, 2243, 2682 (M67), 6253, and 6791; King 2 and 11; Trumpler 5 and 19; Haffner 5 and 10; FSR 0465, 1407, and 1521; Melotte 66; Collinder 261; Gaia 2; and Gulliver 45. 

Three of the clusters---M67, NGC 188, and NGC 6791---are part of the WIYN Open Cluster Study \citep[WOCS,][]{mathieuWIYNOpenCluster2000} and have extensive time-series radial-velocity data for stars on the upper main sequence and brighter (M67: \citealp{gellerStellarRadialVelocities2021}; NGC 188: \citealp{narayanWIYNOpenCluster2026}; NGC 6791: \citealp{tofflemireWIYNOPENCLUSTER2014}). We made 6-dimensional (position, parallax, proper motion, and radial-velocity) membership lists of these clusters using Gaia Data Release 3 \citep[Gaia DR3;][]{GaiaDataRelease2023} following the Gaussian mixture model membership determination procedure in \cite{linckWIYNOpenCluster2024}. Membership lists for the other clusters were derived from the 5-dimensional membership probabilities of \cite{huntImprovingOpenCluster2023}, as that study calculated membership probabilities for stars that are fainter than the faint limits of other Gaia proper-motion membership studies \citep[e.g.,][]{cantat-gaudinPaintingPortraitGalactic2020} but are still on the upper main sequence of some clusters. Comparing our membership lists of M67, NGC 188, and NGC 6791 to the membership probabilities of \citeauthor{huntImprovingOpenCluster2023}, we found the rate of false-positive members among their lists dramatically decreased for stars with membership probability greater than 37\%. We used this probability as the cutoff of membership for every other cluster.

From our study sample, we selected those clusters that meet the following characteristics by inspection:

\begin{itemize}
    \item At least several hundred known members;
    \item Of order 20 or more BSSs; 
    \item Well-defined red clump, MSTO, and upper main sequence;
    \item Low differential reddening.
\end{itemize}

These criteria yielded six clusters, ordered by formation age (Section~\ref{subsec:constraints}): Trumpler 19, M67, Berkeley 39, Berkeley 32, NGC 188, and NGC 6791. Their extinction- and distance-corrected CMDs (Section \ref{subsec:constraints}), along with best-fit isochrones (Section \ref{sec:isochrone}) and BSS regions (Section \ref{sec:bss_region}) are shown in Figure~\ref{fig:cmds}. The cluster characteristics, as derived in the following subsections, and their combined numbers of upper main-sequence and giant stars ($\rm G < G_{MSTO} +2$) stars are listed in Table \ref{tab:clusters}.

\subsection{BSS Identification and Characterization Procedure}\label{subsec:bss_characterisitics}

This study uses the CMD position of a BSS to determine its characteristics using stellar evolution models. We use the properties of their host clusters determined through isochrone fitting as a calibration on reddening, distance, and metallicity of each BSS. In Section~\ref{subsec:constraints}, we first discuss the priors we use for fitting and any other constraints and corrections to data. We then fit isochrones to the non-BSS population of each cluster in Section~\ref{sec:isochrone} and use the cluster isochrone, ZAMS, and TAMS to identify BSSs in Section~\ref{sec:bss_region}. Finally, we discuss our photometric determination of BSS parameters in Section~\ref{subsec:bss_fits}.

As a note, we treat the BSSs as contributing all of the light from each system in Gaia passbands \citep[G, BP, RP,][]{GaiaDataRelease2023}. This is a reasonable assumption for most systems, as a WD companion will typically be significantly fainter in the optical than the BSS and only 3 of the 97 BSSs in the WOCS clusters show a double-line spectroscopic binary, where the secondary light is from a relatively equal-mass main-sequence star (usually mass ratio $q = \frac{M_2}{M_1} > 0.7$). Of these three stars (M67: WOCS 2009, NGC 188: WOCS 5078, NGC 6791: WOCS 54008), we exclude WOCS 2009 (S1082) in M67 as it contains at least three main-sequence stars contributing to the photometry of the system \citep{quitral-pierartMassluminosityAnomaliesPlausible2025}. Finally, WOCS 5885, a BSS on the ZAMS of NGC 188, may have a hot post-asymptotic giant branch companion that contributes a significant amount of visible light to the photometry of the system \citep{subramaniamHotCompanionBlue2016}; we also remove this star from our sample. 

\subsubsection{Reddening, Distance, Metallicity, and Age}\label{subsec:constraints}

Cluster stars were differentially dereddened using the \textsf{dustmaps} Python package \citep{greenDustmapsPythonInterface2018} and the dust map of Schlegel, Finkbeiner, and Davis (SFD; \citeyear{schlegelMapsDustInfrared1998}, recalibrated by \cite{schlaflyMeasuringReddeningSloan2011}). The one exception is Trumpler 19, as applying the differential correction increased the scatter on the main sequence; we instead used the minimum reported reddening among cluster members as the basis for our analysis here. During our isochrone fitting procedure, we allowed for a universal reddening correction for all stars in each cluster using a flat prior truncated at 10\% above and below the value from SFD. 

We used the population mean and standard deviation of distance measurements for upper main-sequence and giant member stars from \cite{bailer-jonesEstimatingDistancesParallaxes2021} to create a Gaussian prior on distance for each cluster. To constrain distance further, we required the luminosity of the red clump region of the isochrone to closely match that of the observed red clump.

The metallicity of each cluster was given a Gaussian prior based on recent spectroscopic measurements: Berkeley 39 \citep[-0.20 $\pm$ 0.04 ][]{bragagliaSearchingMultipleStellar2012}; Berkeley 32 (\citealp[$-0.29 \pm 0.04$]{sestitoOldOpenClusters2006}; \citealp[$-0.30 \pm 0.02$]{carreraChemicalAbundanceAnalysis2011}); NGC 6791 \citep[$0.34 \pm 0.10$ from the combined samples of][]{merchantboesgaardOLDSUPERMETALRICHOPEN2015, carraroNGC6791Exotic2006}; NGC 188 (\citealp[$0.064 \pm 0.018$]{sunWIYNOpenCluster2022}; \citealp[$-0.030\pm0.015$][]{casamiquelaAbundanceAgeRelations2021}; \citealp[$0.09\pm0.02$][]{donorOpenClusterChemical2020}); and M67 (\citealp[$-0.075\pm0.007$]{casamiquelaAbundanceAgeRelations2021}; \citealp[$0.01\pm0.03$][]{donorOpenClusterChemical2020}). To our knowledge, no high-resolution spectroscopic determination of the metallicity of Trumpler 19 has been conducted. Instead, we used the reported metallicities of the cluster giants from Gaia XP spectra \citep{khalatyanTransferringSpectroscopicStellar2024} of $-0.07 \pm 0.10$.

Constraints on the age of each cluster were chosen by finding an upper and lower bound on age that produced an isochrone that roughly fit the shape and location of the cluster by inspection. We used a uniform prior for age within those bounds.

\subsubsection{Isochrone Fitting}\label{sec:isochrone}

\begin{deluxetable*}{cccccccc}\label{tab:clusters}
\tablecaption{Cluster Characteristics}

\tablehead{\colhead{Cluster} & \colhead{Age} & \colhead{Distance} &  \colhead{[Fe/H]} & \colhead{$\rm\langle E(BP-RP)\rangle$} &\colhead{Stars\tablenotemark{a}} & \colhead{BSS} & \colhead{$\rm M_{MSTO}$\tablenotemark{b}} \\ 
\colhead{} & \colhead{[Gyr]} &  \colhead{[pc]} & \colhead{} & \colhead{[mag]} & \colhead{} & \colhead{} & \colhead{[\Msun]}}

\startdata
Trumpler 19  & $4.02$ & $2336$ & $-0.16$ & $0.32$ & 501 & 21 & 1.23 \\[-5pt]
 & \scriptsize $4.21\,[3.85,\,4.50]$ & \scriptsize $2306\,[2239,\,2394]$ & \scriptsize $-0.10\,[-0.14,\,-0.07]$ & \scriptsize $0.31\,[0.29,\,0.32]$ &  &  & \\[5pt]
M67  & $4.10$ & $834$ & $0.05$ & $0.05$ & $470$ & $19$ & $1.29$ \\[-5pt]
 & \scriptsize $4.11\,[3.90,\,4.33]$ & \scriptsize $835\,[822,\,859]$ & \scriptsize $0.07\,[0.05,\,0.09]$ & \scriptsize $0.04\,[0.03,\,0.05]$ &  &  & \\[5pt]
Berkeley 39  & $5.90$ & $4151$ & $-0.20$ & $0.21$ & $737$ & $25$ & $1.10$ \\[-5pt]
 & \scriptsize $5.96\,[5.80,\,6.20]$ & \scriptsize $4151\,[4142,\,4178]$ & \scriptsize $-0.20\,[-0.20,\,-0.19]$ & \scriptsize $0.21\,[0.21,\,0.21]$ &  &  & \\[5pt]
Berkeley 32  & $6.05$ & $3093$ & $-0.34$ & $0.21$ & $474$ & $42$ & $1.06$ \\[-5pt]
 & \scriptsize $5.80\,[5.40,\,6.07]$ & \scriptsize $3095\,[3088,\,3105]$ & \scriptsize $-0.34\,[-0.34,\,-0.33]$ & \scriptsize $0.22\,[0.21,\,0.24]$ &  &  & \\[5pt]
NGC 188  & $6.57$ & $1778$ & $0.00$ & $0.12$ & $549$ & $22$ & $1.11$ \\[-5pt]
 & \scriptsize $6.28\,[5.91,\,6.57]$ & \scriptsize $1794\,[1777,\,1818]$ & \scriptsize $0.00\,[-0.02,\,0.00]$ & \scriptsize $0.13\,[0.12,\,0.14]$ &  &  & \\[5pt]
NGC 6791  & $8.61$ & $4103$ & $0.35$ & $0.17$ & $2672$ & $55$ & $1.11$ \\[-5pt]
 & \scriptsize $8.77\,[8.30,\,9.16]$ & \scriptsize $4088\,[4035,\,4127]$ & \scriptsize $0.33\,[0.30,\,0.36]$ & \scriptsize $0.18\,[0.16,\,0.18]$ &  &  & \\[5pt]
\enddata
\tablecomments{The top row for each cluster gives our best-fit set of parameters, as used in Figure \ref{fig:cmds}. The second row is the median and IQR of our posterior distribution for each parameter. We drew samples from our posterior in a bootstrap process as a calibration for distance, [Fe/H], and E(BP-RP) for our determination of each BSS mass and age, as plotted in Figure \ref{fig:bss_mass_age}.}
\tablenotetext{a}{Number of member stars of brightness $\rm G< G_{MSTO} + 2$}
\tablenotetext{b}{The MSTO mass of our best fit isochrone given for reference. \Mmsto\;was found separately for every draw in our bootstrap.}

\end{deluxetable*}

We developed a Monte Carlo Markov Chain (MCMC) procedure using the Python package \textsf{emcee} \citep{foreman-mackeyEmceeMCMCHammer2013} to fit MIST isochrones \citep[Version 1.2,][]{dotterMESAISOCHRONESLAR2016, choiMESAISOCHRONESSTELLAR2016, paxtonMODULESEXPERIMENTSSTELLAR2011, paxtonMODULESEXPERIMENTSSTELLAR2013,paxtonMODULESEXPERIMENTSSTELLAR2015} to each cluster. The photometry of a MIST isochrone can be described by a unique set of age, metallicity, distance, and reddening. We used the \textsf{isochrones} Python package \citep{mortonIsochronesStellarModel2015} to interpolate an isochrone for each set of these parameters. Each candidate isochrone was fit to the stars of the upper main sequence (G $< \rm{G_{MSTO}+2}$), subgiant branch, and red clump, as the shape of the upper main sequence and subgiant branch of a cluster are very sensitive to cluster age and metallicity, whereas their luminosities are very sensitive to age and distance; likewise, the color and magnitude of the red clump is sensitive to reddening, metallicity, and distance. We did not fit the giant branch as MIST isochrones can exhibit discrepancies in this region due to challenges modeling convective efficiency, opacity treatments, and core helium-burning phases \citep{reyesIsochroneFittingOpen2024}. For our fitting, we removed likely BSSs and photometric binaries by discarding stars that were bluer or much brighter than the MSTO or redder than the dense ridge of stars along the single-star main sequence. 

For each star, we constructed a multivariate Gaussian distribution based on its Gaia G, BP, and RP photometry and photometric error. During every step of the MCMC, for each star we found the point on a given candidate isochrone with photometry that had the maximum log probability from that star's Gaussian distribution. We then summed these values for every star and added the logs of the prior probabilities of the set of parameters used to create that candidate isochrone.

The largest uncertainty in our priors is the age of each cluster, so we started 12 walkers in each 100 Myr division of the age distribution and drew initial values for metallicity, reddening, and distance from the other prior distributions. Walkers were allowed to take steps until the chain converged to a stable posterior distribution. Typically, the first 500 steps of each walker were discarded as burn-in. Depending on the cluster, the final posterior distributions were based on 250,000--1,300,000 draws. The best fit for each cluster was the set of parameters with the maximum log probability, reported in Table \ref{tab:clusters}. 

The posterior distributions are complex, somewhat multi-modal, and show correlations between parameters as expected (e.g., increased reddening decreases metallicity or increased age decreases distance). As such, we also report the median and inter-quartile range (IQR) of the posterior distribution for each parameter in Table \ref{tab:clusters}. These values are not standard deviations nor are they fully independent of each other, so should not be used as standard deviations. However, the median values and range of the IQR can still be compared between clusters. For example, our analysis is unable to distinguish a difference in age for Trumpler 19 and M67, but indicates they are younger than the other clusters.

\subsubsection{BSS Region}\label{sec:bss_region}

We defined the BSS region for each cluster as between the ZAMS and the TAMS and separated cleanly from stars on the main sequence by at least 0.02 mag in color (more than double the median error on color for stars in this study). The ZAMS and TAMS were set by our best-fit isochrone. We also excluded five stars directly above the MSTO of clusters as possible photometric binaries (Trumpler 19: 2, Berkeley 32: 1, NGC 188: 1, NGC 6791: 1). Figure \ref{fig:cmds} shows the BSS region for each cluster and Table \ref{tab:clusters} lists the number of BSSs in each cluster. We identify a total of 184 BSSs across the 6 clusters.

\subsubsection{BSS Fits and Errors}\label{subsec:bss_fits}

\begin{splitdeluxetable*}{cccccccBccccccc} \label{table:bss}
\tabletypesize{\footnotesize}
\tablecaption{BSS Properties}
\tablehead{
   \colhead{Gaia DR3 ID} & \colhead{WOCS ID} & \colhead{Cluster} 
   & \colhead{RA (ICRS)} &  \colhead{Dec (ICRS)} 
   & \colhead{Evolution Age} &  \colhead{Mass} &  \colhead{$T_{eff}$}
   &\colhead{log g}&\colhead{log L}&\colhead{Radius} 
   &\colhead{\age}&  \colhead{\Mrel}  \\
    \colhead{} & \colhead{} &\colhead{} & \colhead{} & \colhead{} & \colhead{Gyr} & \colhead{\Msun}& \colhead{K} & \colhead{} & \colhead{} & \colhead{$\rm{R_\odot}$} &\colhead{}&\colhead{}}
\startdata
$573933322966008960$	& $8104$ & NGC 188	&$10.064616585556992	$&$85.06345774998789	$&${3.03}\,[{2.42},\,{3.61}]$	& ${1.08} \,[{1.07},\,{1.10}]$	& ${6081} \,[{6039},\,{6127}]$	& ${4.38} \,[{4.37},\,{4.39}]$	& ${0.18} \,[{0.17},\,{0.20}]$	& ${1.11} \,[{1.10},\,{1.13}]$	& ${0.41} \,[{0.34},\,{0.47}]$	& ${0.97} \,[{0.96},\,{0.98}]$\\ 
$573937961530665088$	& $5467	$& NGC 188	&$12.60415212043305	$&$85.18392229634908$	&${3.09} \,[{2.59},\,{3.53}]$	& ${1.10} \,[{1.08},\,{1.11}]$	& ${6126} \,[{6085},\,{6171}]$	& ${4.36} \,[{4.35},\,{4.37}]$	& ${0.22} \,[{0.20},\,{0.24}]$	& ${1.15} \,[{1.13},\,{1.16}]$	& ${0.44} \,[{0.39},\,{0.49}]$	& ${0.98} \,[{0.97},\,{0.99}]$\\ 
$573938305128033920$	& $5434$	& NGC 188	&$12.22732779293928$	&$85.21020056315065$	&${2.31} \,[{2.19},\,{2.42}]$	& ${1.33} \,[{1.31},\,{1.34}]$	& ${6573} \,[{6529},\,{6623}]$	& ${4.15} \,[{4.15},\,{4.16}]$	& ${0.63} \,[{0.62},\,{0.65}]$	& ${1.60} \,[{1.58},\,{1.62}]$	& ${0.66} \,[{0.64},\,{0.67}]$	& ${1.19} \,[{1.18},\,{1.19}]$\\
\enddata
\tablecomments{This table is available in machine-readable format. A portion is shown here for guidance regarding its form and content. The first value is the average and the values in the square brackets are the value at the 16th and 84th percentiles, respectively.}
\end{splitdeluxetable*}

To estimate the masses and evolution ages of BSSs, we fit the photometry of each BSS using MIST evolutionary tracks. To determine the uncertainties on BSS properties, we implemented a bootstrapping procedure where we drew from sets of parameters from our posterior sample of isochrone fits based on their probabilities. As noted in Section \ref{sec:isochrone}, there are correlations, degeneracies, and complex distributions among the parameters that define an isochrone. By drawing from our posterior sample, rather than the distributions of the parameters themselves, we are able to guarantee that each set of parameters produces a reasonable fit to the cluster. The parameters of each draw were used to determine the set of possible evolutionary tracks of all BSSs in the cluster in order to treat the entire BSS population in the same way on a given draw. For each draw, we also recorded the mass of stars at the MSTO (\Mmsto) and the ages of stars of each BSS mass at the ZAMS and TAMS as defined by MIST \citep{dotterMESAISOCHRONESLAR2016}. For each cluster, we drew 1,000 samples, which was sufficient for our distributions in age and mass to converge for each BSS. 

Like with other cluster stars, for each BSS we created a multivariate Gaussian using its respective Gaia G, BP, and RP magnitude and photometric error. By using the cluster parameters for distance, metallicity, and reddening, the photometry of any point on an evolutionary track in effect becomes a function of age and mass. To compare stars of many masses and during evolutionary phases of varying duration, MIST implements a metric to delineate equivalent evolutionary phases \citep{dotterMESAISOCHRONESLAR2016}; to sample linearly in both age and mass we normalized for this quantity. We can then measure the probability that a given combination of age and mass produces the photometry of a specific BSS. By examining all age and mass combinations for each iteration of our bootstrap process, we can estimate the probability density function for age and mass of all BSSs accounting for the uncertainties in cluster characteristics. 

Table~\ref{table:bss} gives the results of our analysis for the properties of all 184 BSSs identified in Section~\ref{sec:bss_region} along with their Gaia DR3 ID, WOCS ID (if defined), cluster, right ascension, and declination. We report the values of evolution age, mass, $T_{\rm{eff}}$, log $g$, log $L$, and radius at the probability-weighted average and probability-weighted 16th and 84th percentiles. We further provide these values for the main-sequence relative age (\age) and turnoff-relative mass (\Mrel), defined in Section~\ref{sec:mass_ages}.

We examined the impact systematic errors (e.g., on [Fe/H]) would have on our measurements of \Mrel\ and \age. We found that in general measurements changed by at most low-single digit percentages (1--5\%), except for region of the CMD where BSSs spend much time (i.e., near the ZAMS where evolution age estimates were all low but could differ by many hundreds of millions of years) or where there are degeneracies in evolutionary tracks (i.e., the Henyey hook at the end of the main-sequence, which could also change evolution age estimates by hundreds of millions of years and mass by $\sim0.05$ \Msun). As seen in Section~\ref{sec:mass_ages}, the error bars in these regions are larger but are accounted for in our analyses, suggesting our results in Section~\ref{sec:results} are robust to such errors.

Finally, an indication that our determined masses are accurate is given in Section~\ref{subsec:bss_donors}, where we show that the derived WD companion masses are as expected for core masses of AGB and RGB stars.

\section{Results}\label{sec:results}

\subsection{BSS Masses and Ages}\label{sec:mass_ages}

\begin{figure*}
    \centering
    \includegraphics[width=\linewidth]{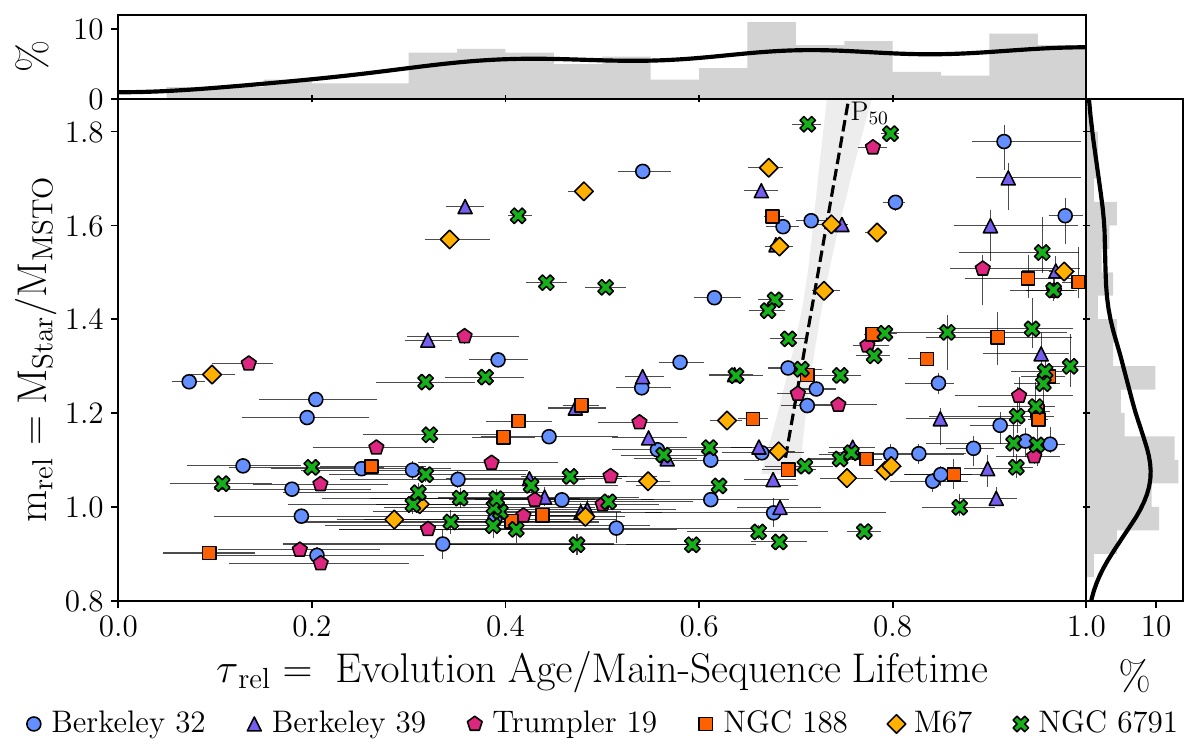}
    \caption{The main panel shows for each BSS what percentage of main-sequence lifetime a single star in its CMD position would have completed (\age) versus the mass of the BSS relative to the turnoff mass of the cluster (\Mrel). The BSSs are plotted at their expected values on each axis with error bars showing the probability-weighted 16th and 84th percentile values. Many of the error distributions are correlated between mass and evolution age or are multi-modal (e.g., around the convective hook near the MSTO); simple error bars are shown for ease of viewing. We plot the median \age\;of the BSSs for a given \Mrel\;derived from our quantile regression as a dashed line along with 1-$\sigma$ errors on the location of the median as the light grey area. The top and side panels display one-dimensional projections of the data as density histograms of the percentage of the population within 5 percentage increments. We also show the KDE of our data, scaled relative to the histograms by dividing the KDE probabilities by the number of bins.}
    \label{fig:bss_mass_age}
\end{figure*}

Based on their CMD positions, the BSSs in our clusters range in mass from 0.96 \Msun\;to 2.25 \Msun\;and evolution ages from 190 Myr (consistent with 0 Myr) to 8.1 Gyr. To directly compare BSSs in clusters of different ages and metallicities, we use relative metrics for their masses and ages.

To estimate the amount of mass gained during an interaction, \cite{jadhavBlueStragglerStars2021} defined the quantity \textit{fractional mass excess} as the difference between the mass of a BSS and the mass of a star at the MSTO normalized by \Mmsto. Although formally this quantity is defined for BSSs with masses below \Mmsto, having a negative excess implies that such a BSS lost mass, whereas here we presume that all BSSs are actually mass gainers. Thus instead we define the quantity \textit{turnoff-relative mass}: 

\[m_{\rm{rel}} = \frac{M_{\rm{BSS}}}{M_{\rm{MSTO}}}\]

In the same vein, to compare the evolution ages of BSSs across masses and metallicities, we use \textit{main-sequence-relative age} to compare the evolution age of a BSS to the total main-sequence lifetime $\tau_{\rm{MS}}$ of a single star of the same mass:

\[\tau_{\rm{rel}} = \frac{\rm{Evolution\; Age}}{\tau_{\rm{MS}}}\]

\noindent Effectively, $\tau_{\rm{rel}}$ is a relative measure of how far evolved is the interior structure of a BSS, from 0 (ZAMS) to 1 (TAMS).  In contrast to equivalent evolutionary phases in MIST \citep{dotterMESAISOCHRONESLAR2016}---which is designed to capture similar stellar structure of stars of many masses, \age\;is linear in time and thus is more useful toward thinking about how populations change over time. 

Because measurements of BSS evolution age, $\tau_{\rm{MS}}$, \Mmsto, and \Mrel\ depend on the age and metallicity of the host cluster, we measured each of these quantities for every combination of age and mass in every draw of our bootstrap process in Section \ref{subsec:bss_fits} and used the probability of each combination to derive errors on these quantities.

Figure~\ref{fig:bss_mass_age} shows \age\;versus \Mrel\;for every BSS along with the probability-weighted 16th and 84th percentile values of each quantity. The top and side panels present one-dimensional projections of the data. We show both density-weighted histograms with bin-widths of 0.05 and Gaussian kernel density estimations (KDE) of our data.\footnote{As BSSs must be within $0\leq$\age$\leq1$ by definition, in order for the KDE of the top panel to integrate to 1, we reflect the KDE at the boundaries using \textsf{kalepy} \citep{kelleyKalepyPythonPackage2021}.}

By inspection, there is a change in the \age\;distribution above and below \Mrel\;$\sim$ 1.4; we use this threshold to divide high- and low-mass BSSs in the analysis below. We find $20 \pm 3 \%$ of BSSs have masses greater than 1.40 times the $\rm M_{MSTO}$, $64 \pm 6 \%$ have masses between $\rm M_{MSTO}$ and 1.40 $\rm M_{MSTO}$, and $16 \pm 3\%$ of BSSs are less massive than $\rm M_{MSTO}$. This matches the observation of \cite{leinerCensusBlueStragglers2021} that approximately 70\% of BSSs in clusters older than 2 Gyr are less than 0.4 \Msun\;above \Mmsto\;(corresponding to \Mrel\;$\sim$ 1.35 for clusters in this study).\footnote{3 of the 6 clusters (M67, Berkeley 39, and NGC 188) examined here are also in \cite{leinerCensusBlueStragglers2021}. The trend of most BSSs being lower mass also holds true with the 3 additional clusters examined here (Berkeley 32, Trumpler 19, and NGC 6791).}

We find that $50 \pm 5 \%$ of all BSSs have the CMD locations of stars that are in the last third of their main-sequence lifetimes, whereas only $15 \pm 3 \%$ are in locations of stars that would be in the first third. This can be seen in the KDE in the top panel, which starts from near zero at \age=0.1 and rises to a plateau beyond \age=0.7, showing a pile-up of BSSs at large \age. A Kolmogorov–Smirnov (K-S) test shows that the distribution of BSSs in \age\;significantly deviates from a uniform distribution (p $= 7.8*10^{-7}$).\footnote{Two-sample K-S tests reject that there are any differences between the \age\ (p$=0.61$) or \Mrel\ (p$=0.84$) distributions of the WOCS clusters and those with membership from \cite{huntImprovingOpenCluster2023}, suggesting there is no membership-related bias in these findings.} 

This pile-up at high \age\;has a dependence on \Mrel. $78 \pm 15 \%$ of the high-mass group (\Mrel\;$\geq1.4$) have the CMD locations of stars in the last third of their main-sequence lifetimes with only one star in a CMD location indicative of a star in the first third of its main-sequence lifetime, showing a significant deviation from a uniform distribution (p $= 2.7 * 10^{-7}$). Those between 1 and 1.4 times \Mmsto\;are more uniformly distributed, but also deviate from uniform (p $= 3.2*10^{-4}$). The distribution of those with \Mrel\;$<1$ formally deviates from uniform (p $=0.047$) with stars clustered at lower \age.

To further investigate this dependence, we performed a quantile regression of the median \age\;for a given \Mrel\;using the python package \textsf{statsmodels} \citep{Skipper_statsmodels_Econometric_and_2010}, also plotted in Figure~\ref{fig:bss_mass_age}. We expect that our data are incomplete at low \Mrel\;for large \age, as some BSSs would occupy the same CMD region as cluster main-sequence stars (i.e., they would be blue lurkers, further discussed below). The lowest \Mrel~that have evolutionary tracks that remain entirely within the BSS region of each cluster as defined in Section~\ref{sec:bss_region} was \Mrel~$\sim$ 1.07, which we adopt for the lower bound of the quantile regression. To estimate errors on the median, we used a Monte Carlo simulation to perform quantile regressions on 10,000 realizations of the BSS \age--\Mrel\;distribution by drawing these properties from the probability distribution functions of each BSS derived in Section~\ref{subsec:bss_fits}. We find that the median \age\;for a given \Mrel\;is above 0.66 for all BSSs with \Mrel\;$>1.07$ (p = 0.0163). Although the median values trend to higher \age\;at higher \Mrel, this effect is not statistically significant. 

As noted, evolutionary tracks of BSSs below roughly 1.07 times \Mmsto\;intersect the main sequences of each cluster. This effect occurred for every cluster but NGC 6791 due to convective hooks of the BSS evolutionary tracks intersecting the main sequence; only NGC 6791 was old enough that the evolutionary tracks for both the low-mass BSSs and the cluster stars do not show convective hooks. This may explain the absence of stars in the bottom right of Figure~\ref{fig:bss_mass_age}: the lower the mass of a BSS, the more its evolutionary track leaves the BSS region, in effect making them blue lurkers. If the distribution in Figure \ref{fig:bss_mass_age} of BSSs slightly below 1.07 times \Mmsto~is the same as those slightly above, this may suggest that there are several tens of blue lurkers among the upper main sequences of these clusters. In M67, \cite{leinerBlueLurkersHidden2019} identified main-sequence binary interaction products through rapid rotation; six of these stars are blue lurkers using our membership lists and BSS region. To first order, if the simple ratio of 6 blue lurkers to the 19 BSSs of M67 holds for other clusters, this number of blue lurkers is plausible.

In this subsection, we have shown that BSSs are preferentially found nearer the end of their main-sequence lives and that few high-mass BSSs are near the ZAMS. Such an increase of the number of BSSs with increasing \age\ indicates that BSSs must form at later \age\ rather than all start on the ZAMS. This in turn suggests that many BSSs form with evolved cores having helium abundances larger than primordial. In the following sections, we explore this idea as a possible explanation of this trend.

\subsection{BSS Core Composition at Transformation}\label{subsec:core}

\begin{figure*}[ht!]
    \centering
    \includegraphics[width=\linewidth]{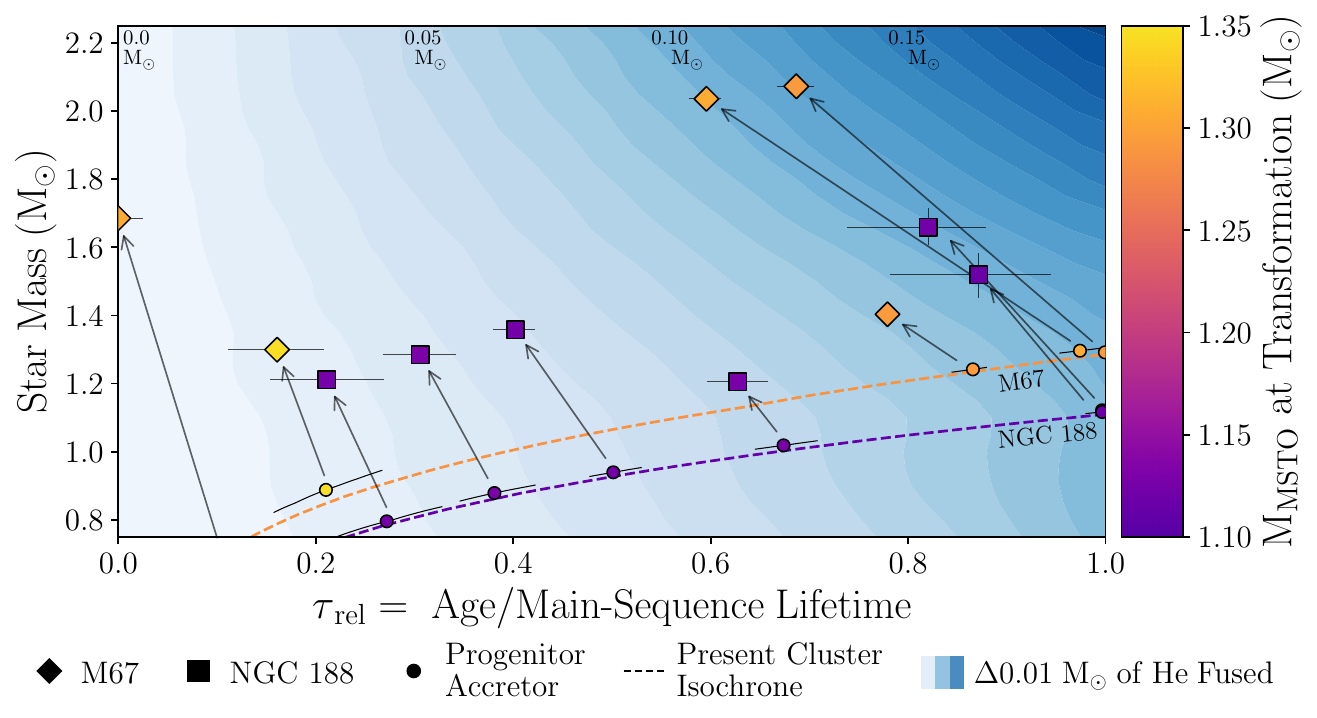}
    \caption{Transformation ages are known for those BSSs that have WD companions with measured cooling ages. We show the main-sequence-relative age (\age) and mass of the 11 BSSs with known WD companions in M67 and NGC 188 (diamond and square markers, respectively) at the time they transformed into BSSs (evolution age - WD cooling age). Markers are color-coded by the \Mmsto\;of the host cluster at the time each star transformed. As these stars have hot WD companions, they are presumed to have undergone mass transfer (see Section~\ref{subsec:bss_donors}). For a mass-transfer BSS with a known transformation age, we can deduce a narrow mass range of the progenitor accretor prior to the interaction. During mass transfer, the amount of helium in the accretor core does not change to first order (although the abundance may change), as this process occurs over a much shorter time-span than the star's main-sequence lifetime. The blue background areas show equivalent amounts of helium fused in a main-sequence star of a given mass (y-axis) as a function of \age. Regions are broken into increasing increments of 0.01 \Msun\;and several are labeled with the amount of helium fused for ease of comparison. Because we know the cluster age at transformation, we also know the evolutionary state of every main-sequence star of the cluster at that time using an isochrone. We plot the current-day isochrones of M67 and NGC 188 for reference as dashed lines. Finally, we can find the portions of an isochrone that have an equivalent amount of helium created as the BSS CMD position indicates that it has after transformation and determine the mass and \age\;of its progenitor accretor (plotted as small circles with the same color-scale as the BSS).}
    \label{fig:he_content}
\end{figure*}

11 of the BSSs in M67 and NGC 188 have observed WD companions with measured masses and cooling ages from either UV spectroscopy or UV photometry \citep{gosnellIMPLICATIONSFORMATIONBLUE2015, sindhuUVITOpenCluster2019,jadhavUVITOpenCluster2019, gosnellConstrainingMasstransferHistories2019,leinerObservationsSpindownPostmasstransfer2018, vernekarPhotometricVariabilityBlue2023, nineDetectionBLM672023, palDiscoveryBariumBlue2024}. The WD masses are consistent with dynamical minimum masses from orbit solutions (see Section~\ref{subsec:bss_donors}). The cooling ages are a metric of the transformation age of a BSS as mass transfer between an accretor and donor ends when the WD is exposed. The observed cooling ages are a few hundred Myr or less. \cite{gosnellIMPLICATIONSFORMATIONBLUE2015} observe that the BSSs with known WD companions in NGC 188 are found both on and well off the ZAMS, demonstrating that distance from ZAMS need not correspond to BSS (transformation) age.

Given the CMD position and transformation age of a BSS, the \age\ of the BSS at transformation can be derived from current evolution age minus the transformation age. In Figure~\ref{fig:he_content}, we show the BSSs with known WD companions at the \age\ of their transformations. Evidently BSSs can form at all \age. Notably, however, only one of the eleven BSSs was on the ZAMS (\age\;$=0$) after it transformed. On the other hand, more than half of these BSSs had \age\;at transformation $\geq 0.6$. 

This distribution of formation \age\ is consistent with our interpretation of Figure~\ref{fig:bss_mass_age} that the rising number of BSSs with \age~in all clusters indicates that BSSs must be joining this population at higher \age. \cite{mathieuBlueStragglersFriends2025} argue that the NGC 188 BSSs with known WD companions that are not on the ZAMS imply these BSSs formed with already evolved cores. We explore this idea next.

Broadly speaking, the CMD location of a star is determined by the properties of its core, the conditions setting opacity in its envelope, and its rate of rotation. For BSSs, a key question is how accreted mass impacts the ongoing interior evolution of the star and how that could affects observable properties. 

Recent stellar evolutionary models of mass-transfer products show that as mass is accreted the accretor structure adjusts to this higher mass by growing its core with material that was originally just outside of the core \citep[e.g.,][discussed in more detail in Section~\ref{subsec:BSS_structure}]{schneiderPresupernovaEvolutionFinal2024, waggAsteroseismicImprintsMass2024, sunWOCS5379Detailed2021}. As the new core material is hydrogen-rich, it lowers the core helium abundance, rejuvenating the star to behave like a higher-mass star at a younger evolutionary age, which then continues to evolve like a single star that has not undergone an interaction. In detail, physical effects, such as convection or chemical gradients, can impact the amount of hydrogen brought into regions of the core undergoing the highest rates of fusion \citep{schneiderPresupernovaEvolutionFinal2024}. 

In order to take advantage of the observable effects of helium evolution in cores, we used Modules for Experiments in Stellar Astrophysics \citep[MESA; version 24.03.01][]{paxtonMODULESEXPERIMENTSSTELLAR2011, paxtonMODULESEXPERIMENTSSTELLAR2013, paxtonMODULESEXPERIMENTSSTELLAR2015,paxtonModulesExperimentsStellar2018,paxtonModulesExperimentsStellar2019, jermynModulesExperimentsStellar2023} to model the main-sequence evolution of solar-metallicity single stars between 0.7 and 2.5 \Msun, spanning the range of progenitor accretors and BSS masses in the old open clusters. Most model parameters were the default values of MESA, however, we used the hydrogen and helium abundances of M67 from \cite{reyesIsochroneFittingOpen2024} and the predictive mixing algorithm with the Ledoux criterion from \cite{paxtonModulesExperimentsStellar2018} to more accurately model convective core masses.\footnote{
The inlist is available on Zenodo: \\\dataset[doi: 10.5281/zenodo.20071909]{\doi{10.5281/zenodo.20071909}}} We determined the amount of helium produced in the cores of these stars throughout core hydrogen burning and show this in the background of Figure~\ref{fig:he_content}. Each shaded region of Figure~\ref{fig:he_content} represents equal amounts of helium (in increments of 0.01 \Msun) produced by main-sequence stars of differing masses (y-axis) versus time, shown as the percentage of each star's main-sequence lifetime (i.e., \age). 

During the time that a star accretes material from the envelope of the donor, the amount of core helium will not change (at least not to first order as the time-span of mass transfer is much shorter than the star's main-sequence lifetime), although the fractional helium abundance may change. Assuming that the CMD position of a BSS is reasonably representative of its core evolutionary state, as seen in the models discussed in Section~\ref{subsec:BSS_structure}, this means we can map the mass and \age\;at transformation of a BSS to a specific mass and evolutionary state of an accretor based on the amount of helium that has been created on the main sequence during core hydrogen burning. To do so, we use our MESA models to interpolate the amount of helium created by a star of the BSS mass at its transformation \age\;and the amount of helium created by all main-sequence stars in the cluster at that transformation age from MIST isochrones. We can then identify a narrow range of progenitor accretor masses and evolutionary states that could have created that BSS. We show the location of the progenitor accretor for each BSS in Figure~\ref{fig:he_content} along with the mapping of the error of the transformation age. For reference, we further show the current-day positions of main-sequence stars for both clusters as isochrones.

The most evolved BSSs (\age\ $\gtrsim$ 0.6) have the same amount of helium in their cores as stars near the MSTO at the time they transformed into BSSs. More generally, the amount of helium created by a star at the MSTO of an old cluster is the same amount of helium that a star 0.5--0.8 \Msun\ more massive than the MSTO would have created by 60--70\% of its main-sequence lifetime. For these old clusters, such stars corresponds to BSSs of \Mrel\ $\geq1.4$. 

The lower-mass BSSs ($m<1.4$ \Msun) span a range of \age\;at transformation, which indicates that their progenitor accretors had a range of helium created in their cores prior to BSS transformation. Thus they also must have had a range of accretor masses lower than \Mmsto\ (roughly 0.6--1.25 \Msun\ depending on the star). These BSSs are also only 0.15--0.42 \Msun-more massive than their accretors, which suggests they did not undergo conservative mass transfer. 

From MIST, at the time of transformation for these stars, the giants are only a few hundredths of a solar mass more massive than \Mmsto (and in some cases in old open clusters, the same mass due to mass loss on the giant branch). Given the range of accretor masses among these stars, the initial mass ratio ($q=\rm \frac{M_{Accretor}}{M_{Donor}}$) was at least 0.6, with some being near equal-mass binaries. $q$ closer to unity is more likely to lead to stable mass transfer \citep{ivanovaBinaryEvolutionRoche2015}, which in turn may explain how these stars were able to accrete as much material needed to stand out as BSSs. We will explore the initial $q$ of other BSSs more in the next sub-section.

Only one star is on the ZAMS (WOCS 4006 in M67), implying that it formed with only primordial helium in its core. This star has gained a significant amount of mass (more than doubling in mass based on Figure~\ref{fig:he_content}), but does not have a known formation mechanism (see Appendix~\ref{app:categorization}).

We note that three of the four most massive BSSs have helium amounts that are up to 10\% larger than could have been produced by their accretors. This could be a result of different physics in the MIST models used to determine BSS properties and the MESA models run here, or because of physical differences between single-star and interaction-product models, such as chemical gradients at later \age\;inhibiting new hydrogen from reaching the core \citep{schneiderPresupernovaEvolutionFinal2024}. 

In this subsection, we have shown that the BSSs with known WD companions generally did not form on the ZAMS, but rather form at all evolution ages. We further demonstrated that the core helium amounts of the highest mass BSSs at \age $\sim 0.6$ are equivalent to the core helium amounts in stars near the MSTO at the time they formed, indicating that MSTO stars are the progenitor accretors of high-mass BSSs. Lower mass BSSs formed at all evolution ages. This indicates that the accretors forming these stars had a wide range of degrees of evolution on the main sequence prior to interaction, but generally were of $q>0.6$ with many near equal mass. The distribution of BSSs with known WD companions is consistent with the observation in the previous subsection that most BSSs are near the end of their main-sequence lifetimes, as BSSs will continually form at all evolution ages, so as previous generations of BSSs age, they will show a pile-up effect on the CMD. This further means that low-mass BSSs at late evolutionary ages on Figure~\ref{fig:bss_mass_age} may have very different accretor progenitors but appear to be in the same place on the CMD. Whether newly formed BSSs of late evolutionary ages have high or low masses must then be set how much material was accreted, which is a function of mass-transfer efficiency and donor envelope mass, which we discuss in the next subsection.  

\subsection{BSS Mass-Transfer Efficiencies and Progenitor Accretor Masses}\label{subsec:mt_eff_q}

\begin{figure*}[ht!]
    \centering
    \includegraphics[width=\linewidth]{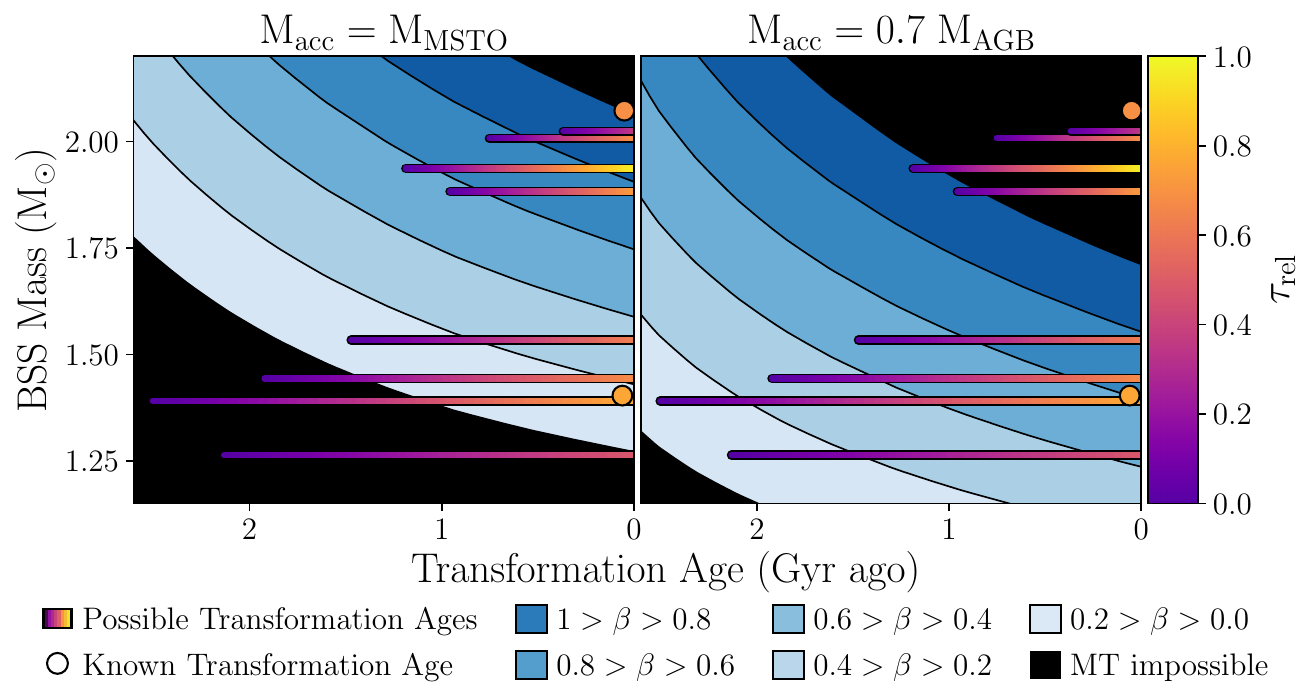}
    \caption{The mass of a BSS is the sum of the mass of its progenitor accretor and the amount of the donor envelope it accretes. For AGB-mass-transfer products, constraints can be placed on the masses of both the donor and accretor as a function of mass ratio ($q$), mass-transfer efficiency ($\beta$), and transformation age, as the mass of the donor AGB (and the mass of its envelope) maps directly to cluster age. The figure shows the combinations of $q$ and $\beta$ that could have created the AGB mass-transfer products in M67 as a function of permitted transformation age. (The same for NGC 188 is shown in Figure~\ref{fig:n188_agb}.) The range of possible transformation ages for a given BSS is shown by a horizontal line plotted at the mass of the BSS, color-coded by what percentage of its total main-sequence lifetime (\age) that transformation age represents. Stars with hot WD companions and known transformation ages are plotted as circles. The left panel shows accretion onto a \Mmsto\ star, and the right panel onto a star of mass $q=0.7\,\rm M_{AGB}$. The background curved blue areas show the fraction of the AGB envelope that must be transferred to the accretor to reach a BSS mass. Areas in black cannot be created by any combination of AGB envelopes and accretor masses, as either there is not enough mass in the envelope and accretor to sum to the mass of the BSS or the BSS is lower mass than the accretor star in that panel at that transformation age.}
    \label{fig:m67_agb}
\end{figure*}

\begin{figure*}
    \centering
    \includegraphics[width=\linewidth]{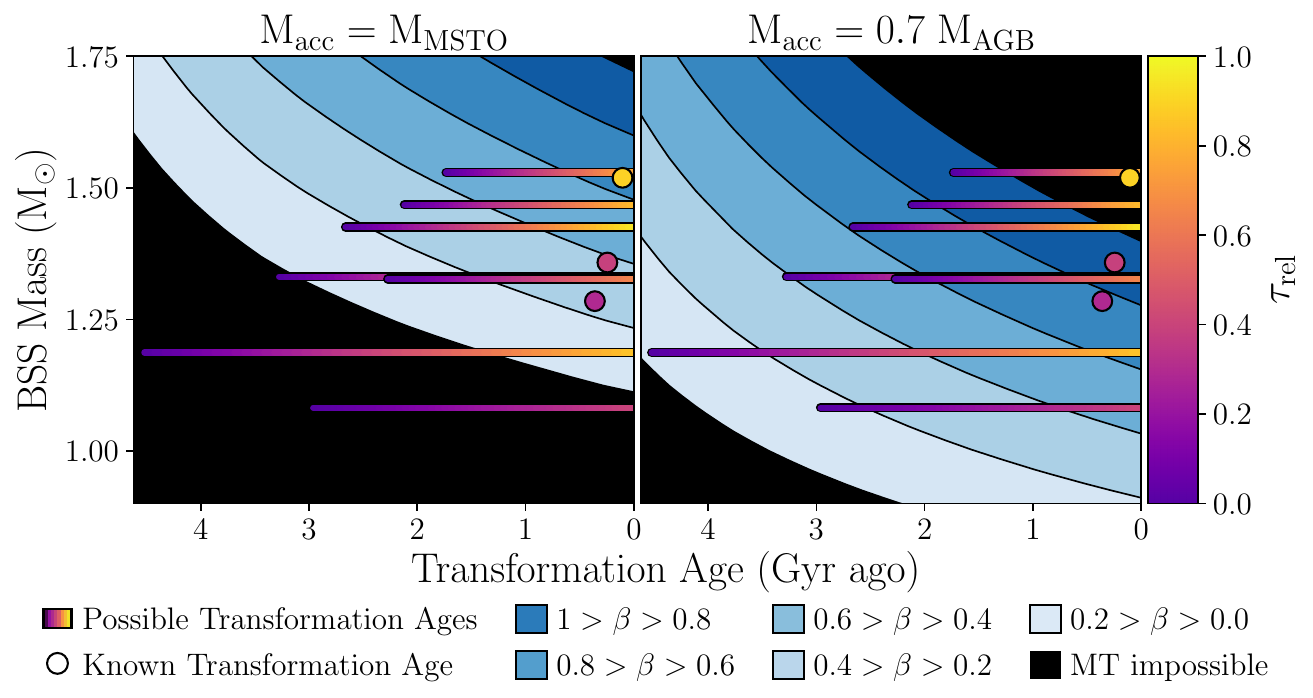}
    \caption{Same as Figure~\ref{fig:m67_agb} but for the AGB mass-transfer products of NGC 188.}
    \label{fig:n188_agb}
\end{figure*}

One of the major unknown properties of a BSS are the masses of its donor and accretor progenitor stars. In the previous subsection, we determined the accretor masses of BSSs with known WD companions. Generally, however, we do not know the masses of either progenitor. Even so, we do know that the accretor mass plus the mass of material accreted sum to the current mass of the BSS. Thus limits can be placed on both the donor and accretor masses if we know the formation mechanism that created the BSS (see Section~\ref{subsec:bss_donors}). 

Asymptotic giant branch (AGB) mass-transfer products in particular offer the strongest constraints on donor and accretor masses. Because the mass of an AGB star corresponds to the age of its host cluster, for AGB mass-transfer products we can describe the final mass of a BSS as a function of cluster age at transformation, $q$, and mass-transfer efficiency ($\beta$) of the AGB star envelope. Despite most BSSs having an unknown transformation age, we can bound this time by the evolution age of each BSS---i.e., the time since a star of that mass would have been on the ZAMS.

Based on MIST evolutionary tracks, the donor AGB star has the smallest envelope to that point in the donor evolution---and thus the smallest amount of available mass to transfer---as the core has grown to $\sim0.5$ \Msun~and the star has lost $\sim0.04$\;\Msun\;from winds by the end of He-core fusion in the red clump. Using MIST, we calculated the envelope masses of AGB stars at the onset of thermal pulses at all cluster ages. The first thermal pulse is the first time that these AGB stars would exceed the radii they had at the tip of the RGB, and thus would undergo mass transfer at or after this point \citep{ivanovaBinaryEvolutionRoche2015}. For reference, the current envelope mass of AGB stars in M67 is $\sim0.8$ \Msun\;and in NCG 188 is $\sim0.6$ \Msun\;in NGC 188.

At any given cluster age we know the masses of potential accretor stars that could form a BSS (those at the MSTO and lower mass). By constraining the donor to be an AGB star, we can then describe the unknown mass of the accretor by combinations of $q$ and transformation age (which determines the mass of the AGB star). For a given $q$ and transformation age, a unique $\beta$ is required to create a BSS of a specific mass.

In Section~\ref{subsec:bss_donors}, we identify 20 BSSs in M67 and NGC 188 that formed through AGB mass-transfer. By combining the donor envelope and potential accretor masses at any given cluster age, we can compare the mass of each BSS (Section~\ref{sec:mass_ages}) to any combination of $q$ and $\beta$ that could have occurred since the earliest time an accretor could have transformed into the BSS. 

In Figure~\ref{fig:m67_agb}, for every AGB-mass-transfer BSSs in M67, we show the fraction of the AGB donor envelope at a specific transformation age that would need to be added to the mass of an accretor to reach the BSS mass.

The left and right panels show the fraction of AGB donor envelope that must be transferred on to an accretor of mass \Mmsto\,or 0.7 $\rm M_{AGB}$, respectively, at the transformation age on the x-axis to produce a given BSS mass (y-axis). We further plot the masses and ranges of possible transformation ages of each AGB-mass-transfer BSS in M67. Figure~\ref{fig:n188_agb} shows the same but for NGC 188. For the five AGB-mass-transfer stars that have WD companions with known cooling ages in the two clusters, we can identify the exact AGB mass and \Mmsto\;at the time of transformation, placing strong constraints on all possible $q$ and $\beta$ for these systems. 

Considering Figure~\ref{fig:m67_agb}, some of the accretors are both massive and have formed recently enough to require that both mass transfer be nearly or entirely conservative (i.e., the entire envelope was transferred or $\beta\sim1$) and the original accretor mass to be near \Mmsto\;(i.e., an evolved main-sequence star at that cluster age). None of AGB-mass-transfer products in NGC 188 (Figure~\ref{fig:n188_agb}) are quite as massive but the higher-mass BSSs with a known WD companion suggests $\beta$ to be at least 0.65. 

Conversely, for the lowest-mass BSSs to have undergone conservative mass transfer, the accretor mass would need to be quite low ($q\sim0.4\;\rm M_{AGB}$, not pictured in the figures), which may raise concerns of plausibility, as unstable (and thus not conservative) mass transfer is more likely to occur at lower $q$ \citep{ivanovaBinaryEvolutionRoche2015}. Instead, these stars more likely underwent non-conservative mass transfer ($\beta<0.5$) on to stars less massive but nearer \Mmsto. For intermediate-mass BSSs, there is a degeneracy between $q$ and $\beta$ that cannot be broken with this analysis, except those with WD companions. 

The five BSSs that have detected hot WD companions that define their transformation age (circles on both the M67 and NGC 188 plots) show the original accretor to be at least $60-70\%$ of the mass of the AGB donor (see also Figure~\ref{fig:he_content}), otherwise the current masses of the BSSs could not be reached with the amount of mass available in the envelope of the AGB donor. This in turn suggests that the minimum $\beta$ for these stars was $\sim0.2-0.4$. However, for the lowest-mass BSSs, even if $q\sim0.7$, $\beta$ had to be 0.6 or lower. This also aligns with Figure~\ref{fig:he_content} that shows low-mass BSSs gaining $0.15-0.4$ \Msun-worth of mass, which corresponds to $0.2\lesssim\beta\lesssim0.6$ of the donor envelope masses. 

We note that the two AGB mass-transfer products in M67 with known WD companions (WOCS 2013 and 9005) are both barium-enriched and have cooling ages of roughly 50-60 Myr \citep{pandeyUOCSVIUVIT2021,palDiscoveryBariumBlue2024, nineWIYNOpenCluster2024}. Because of the identical cooling ages, their donors were both $1.32 \pm 0.02$ \Msun\;AGB stars, which would have envelope masses of $\sim0.8\;\rm M_\odot$. Although both WOCS 2013 and 9005 have similar donor masses, they have very different CMD BSS masses---$2.07\pm0.03$ and $1.40\pm0.01$~\Msun, respectively. \cite{palDiscoveryBariumBlue2024} found that the surface barium abundance of WOCS 9005 suggests 0.15\;\Msun-worth of mass was transferred from its AGB companion, which would be 20\% of the available envelope, in good agreement with the CMD analysis here (see the left panel of Figure~\ref{fig:m67_agb}). There is an indication that the two BSSs have different [Ba/Fe]---1.18 \citep[with a range between 0.82 and 1.55]{nineWIYNOpenCluster2024} and $0.75\pm0.08$ \citep{palDiscoveryBariumBlue2024}, respectively, which may indicate that the two stars accreted different amounts of mass, as seen in the figure. \cite{csehBariumStarsTracers2022} note that predictions of the dredge-up of s-process elements that could be transferred by an AGB star is model dependent and that the lowest-mass star in these models that could dredge-up s-process elements (1.3 \Msun) has to be more metal-poor than M67. Given their WD companion cooling ages, WOCS 2013 and 9005 suggest that solar-metallicity AGB stars of 1.3 \Msun\ will in fact produce and dredge up barium. 

The observations and analyses in this and the previous subsection directly establish an AGB-mass-transfer formation pathway in which more massive BSSs are formed by conservative mass transfer onto helium-enriched MSTO stars (i.e., in nearly equal-mass binaries). Further, low-mass BSSs (and possibly blue lurkers) are formed by non-conservative mass transfer from AGB stars onto lower-mass main-sequence stars. BSSs formed from the lowest-mass main-sequence stars will begin their BSS evolution on the ZAMS, as such accretors have little helium enrichment. These conclusions apply only to BSSs formed by AGB donor stars, but, as we will find in Section~\ref{subsec:bss_donors}, AGB mass-transfer appears to form the plurality, if not the majority, of BSSs in these clusters.

\subsection{BSS Progenitor Donors}\label{subsec:bss_donors}

\begin{figure*}
    \centering
    \includegraphics[width=\linewidth]{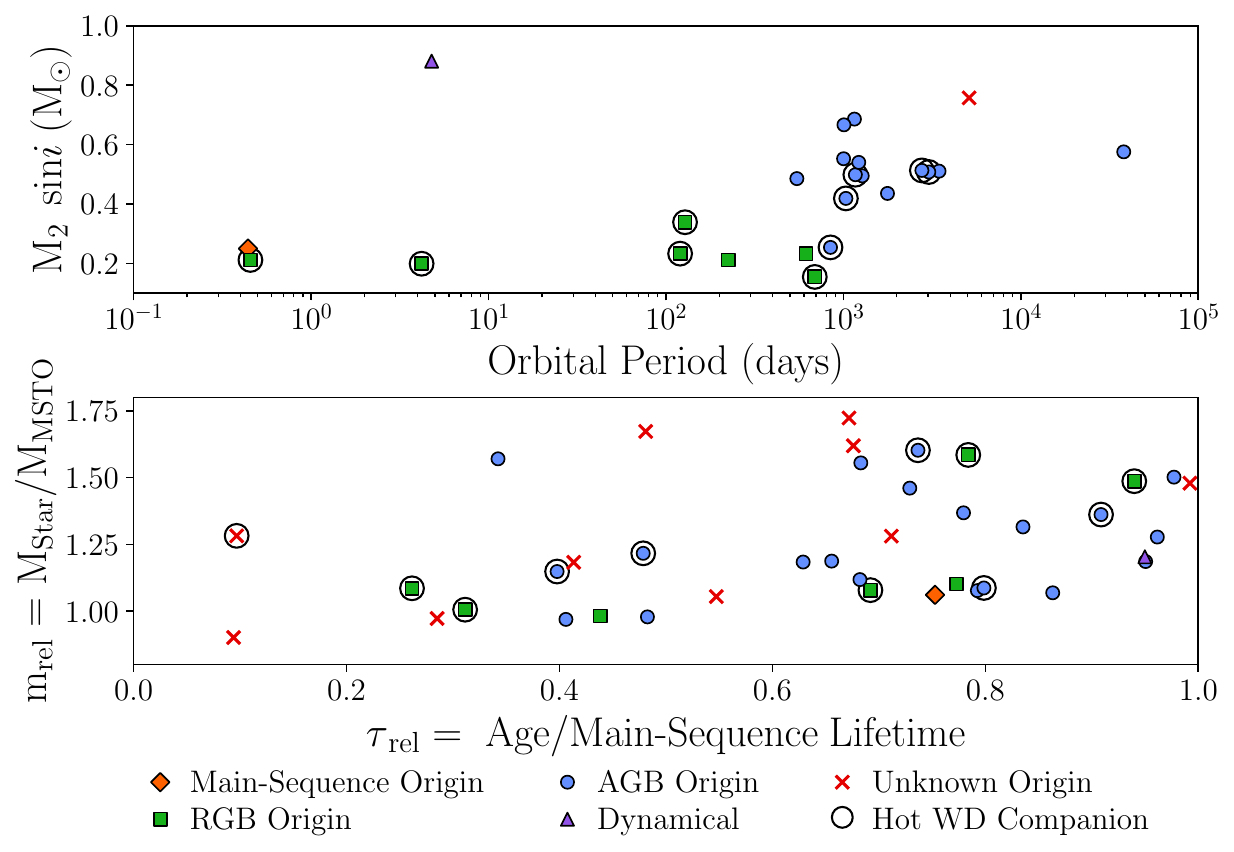}
    \caption{The upper panel shows the period and minimum secondary mass of all BSSs in M67 and NGC 188 with orbit solutions (except the multiple system WOCS 2009 (S1082) in M67), using the BSS mass derived in this work for the primary mass. Using the periods, minimum secondary masses, known WD companions, and abundances, we categorize each star by its formation mechanism (main-sequence origin: orange diamond; RGB origin: green square; AGB origin: blue circle; dynamical origin: purple triangle; unknown origin: red x). Stars with known WD companions are marked with an additional ring. The lower panel shows the distribution of BSSs in these clusters plotted on the same axes as Fig~\ref{fig:bss_mass_age}.}\label{fig:bss_m67_n188}
\end{figure*}

Long-term radial-velocity monitoring has yielded orbit solutions for 25 of 41 BSSs in M67 and NGC 188 \citep{lathamSpectroscopicBinariesM671996, miloneBlueStraggler1901992,gellerStellarRadialVelocities2021, gellerWIYNOPENCLUSTER2009, narayanWIYNOpenCluster2026} and permits limits on orbital periods from radial-velocity variability for many of the remaining BSSs. We use these extensive radial-velocity data to observationally identify formation pathways for many of the BSSs in these two clusters. Specifically, we choose to use donor evolutionary phase (main sequence, RGB, and AGB) as the primary criterion for categorization, as in the most frequent formation channel of mass transfer the combination of orbital period, secondary mass, and abundances provide direct insight into the evolutionary state of the progenitor primary just prior to the interaction. In Appendix~\ref{app:categorization}, we analyze the available information for all 41 BSSs star-by-star to categorize the formation mechanism of each. 

Figure~\ref{fig:bss_m67_n188} shows these categorizations. Altogether, among the 41 BSSs of M67 and NGC 188, at least 20 are of AGB origin, 7 of RGB origin, 1 of main-sequence origin, and 2 of dynamical origin (dynamical exchange with a system that contains a BSS). 11 stars do not have sufficient information to categorize their progenitor donors evolutionary states and formation mechanism, but likely they include mergers (2 candidates), collisions, and mass-transfer products. These results suggest that at least two-thirds of BSSs in old open clusters are mass-transfer products, with (at least) half having undergone AGB mass transfer. Using the number of WD companions they identified among the BSS of NGC 188, \cite{gosnellIMPLICATIONSFORMATIONBLUE2015} predicted that at least two-thirds of BSSs should have WD companions, indicating they had undergone mass transfer. This prediction is in very good agreement with the rate found in this work based on orbit information. This further means that at most $\sim30\%$ of BSSs in these open clusters were formed by mergers and collisions. These results are also in good agreement with and expands upon those of \cite{nineWIYNOpenCluster2024}, who found that Ba abundances among BSSs (both with long orbits and that were non-velocity variable) in somewhat younger clusters (and M67) indicates that AGB mass transfer is the dominant formation mechanism. 

In the upper panel of Figure~\ref{fig:bss_m67_n188}, we show all of the M67 and NGC 188 BSSs with orbit solutions, plotting minimum secondary mass against orbital period. 4 BSSs have periods less than 10 days (1 main-sequence mass-transfer BSS, 2 RGB mass-transfer BSSs, and 1 dynamical-origin BSS); no BSSs have periods between 10 and 100 days; 5 RGB mass-transfer BSSs have periods between 100 and 700 days, and the 14 AGB-origin stars with known orbits have periods greater than 500 days (median of 1195 days). 

There is a marked increase in minimum secondary mass for orbital periods greater than 1000 days, consistent with the smaller cores of RGB donors and the larger cores of the AGB donors. Quantitatively, the AGB-origin population shows a tight clustering around 0.55 \Msun---the core mass of AGB stars with masses appropriate to these clusters---with the small scatter in mass explainable by inclination angle. Because the companions of these long-period systems are all the mass of CO WDs, rather than a random scattering of mass, this indicates the companions are all WD stars, as also found earlier in NGC 188 by \cite{geller_mass_2011}. The RGB mass-transfer population shows a wider scatter in minimum secondary mass ($0.15<m_2\;\sin i<0.33$ \Msun) at masses appropriate to the growing degenerate cores of RGB stars in this cluster, although 5 of 7 are clustered at $0.21\pm0.02$ \Msun. Moreover, the mass at the median inclination angle of $60^\circ$ is only about 0.04 \Msun~larger than the minimum mass for these low masses, implying that at least half of the minimum masses are close to the actual secondary masses. 

That only one star in our sample shows a secondary minimum mass greater than the WD masses appropriate to the mass-transfer pathway of its orbital period indicates that mergers and collisions are not the formation mechanism of most of these stars, as one would expect a broad distribution of companion masses indicative of having main-sequence companions. 

In Figure~\ref{fig:bss_m67_n188}, we also mark which stars have known WD companions (see Section~\ref{subsec:core} for details). Most other BSSs in these clusters have been investigated for WD companions by Hubble and UVIT observations, so stars without a known WD are likely to be older than cooling ages of roughly a few hundred Myr. The frequency of hot WD companions (those that can be seen in UV studies) seems to be higher among BSSs of RGB origin (5 of 7 in RGB versus 5 of 19 AGB). Two of these WDs have directly measured masses \citep[WOCS 4540: 0.52 \Msun, WOCS 5379: 0.42 \Msun;][]{gosnellConstrainingMasstransferHistories2019}; our minimum secondary masses of these stars (0.51 \Msun~and 0.23 \Msun, respectively) are consistent with those measurements, suggesting the BSS masses derived in Section~\ref{subsec:bss_fits} are reasonable. 

The lower panel of Figure~\ref{fig:bss_m67_n188} shows how each formation mechanism population maps into the axes of Figure~\ref{fig:bss_mass_age}. Like with the analysis of all 6 old open clusters in Section~\ref{sec:mass_ages}, the most massive BSSs also show a pile-up at \age\;$>0.6$. Notably, we can see that this effect appears to be present for both RGB- and AGB-origin products and that the distribution of RGB- and AGB-origin stars is fairly uniform in \age\;at lower \Mrel. Both the RGB- and AGB-origin stars have a relatively uniform distribution of WD companions at every \age, which indicates that the RGB stars also follow the same pattern as the AGB-origin stars discussed above: mass transfer onto accretors near \Mmsto\ will start at later \age\ whereas accretion onto the lowest-mass main-sequence stars will begin their BSS evolution nearer the ZAMS. We do also note one AGB-origin BSS at \age $<$ 0.4 and \Mrel $>$ 1.5. Under the model proposed in this work, following the contours of Figure~\ref{fig:he_content}, this star would have undergone fully conservative mass transfer onto a main-sequence star roughly 0.1 \Msun\;below \Mmsto.

Moreover, the most massive BSSs require relatively conservative mass transfer to form. Two of the high \age\ and \Mrel\ BSSs with WD companions are of RGB origin, indicating they may have underwent relatively conservative mass transfer. However, as RGB stars can have several tenths of \Msun\ more envelope than AGB stars, it is possible that mass transfer is less conservative for these stars. Like with the AGB-origin BSS, the low mass RGB-origin BSS could have undergone non-conservative mass transfer. Of particular note, the WD companion of WOCS 5379 (\age=0.26 and \Mrel = 1.08) in NGC 188 has a measured mass of 0.42 \Msun\ indicating a helium WD and RGB mass transfer \citep{gosnellConstrainingMasstransferHistories2019}. \cite{sunWOCS5379Detailed2021} were only able to reproduce this BSS with accretion onto a 1 \Msun\;star with a total mass-transfer efficiency of 22\%. More examples of this mechanism are needed to establish how conservative mass transfer can be in such systems.
 
The current data are too few to make a similar assessments for main-sequence mass transfer, mergers, or collisions, although we do note that the stars of unknown origin, which would include mergers and collisions, are found at all \age\ and \Mrel, including both the highest and lowest \age. 

In this section, we have found that the orbits, minimum secondary masses, and abundances of BSSs demonstrate that mass transfer from AGB and RGB stars is the dominant formation mechanism in old open clusters and that AGB mass transfer accounts for at least half of BSSs. We further have found that RGB-origin BSSs follow roughly the same distribution in \age\ and \Mrel\ as AGB-origin BSSs, which suggests that they undergo the same formation patterns as AGB stars.

\section{Discussion}\label{sec:discussion}
\subsection{Comparison to Other CMD Studies of BSS}

As noted in Section~\ref{sec:introduction}, numerous studies have used CMD locations of BSSs in open clusters to examine the properties and evolution of BSS populations. Here, we discuss our results in the context of three recent Gaia-based studies that also focused on the ages and masses of BSSs. 

\cite{leinerCensusBlueStragglers2021} examined the BSSs of 16 clusters older than 1 Gyr. They find that the BSSs in intermediate age-to-old open clusters ($>$2 Gyr) show a gap on the CMD between bright BSSs and faint BSSs, corresponding to BSS masses $>0.6$ \Msun\;and $<0.4$ \Msun\;above \Mmsto, respectively. This gap shrinks for clusters older than 4 Gyr. We also see a gap in M67 between the high-mass and low-mass AGB mass-transfer products (Figure~\ref{fig:m67_agb}). However, such a gap is not evident for NGC 188 (Figure~\ref{fig:n188_agb}). \cite{leinerCensusBlueStragglers2021} propose that the observed gap may be caused either by conservative versus non-conservative mass transfer or from different mass-transfer formation pathways (e.g., RGB versus AGB mass transfer). As seen in Figure~\ref{fig:bss_m67_n188}, both RGB and AGB mass-transfer products are found among the low- and high-mass BSSs, indicating that formation mechanism likely is not responsible for the gap. 

As noted in Section~\ref{subsec:mt_eff_q}, the most massive AGB mass-transfer products require conservative mass transfer onto accretors of mass $\sim$\Mmsto, whereas the lowest-mass AGB mass-transfer products likely require non-conservative mass transfer onto lower-mass main-sequence stars, in line with the proposal of \citealp{leinerCensusBlueStragglers2021}. An explanation for the gap disappearing by the age of NGC 188 is that there is less mass available for transfer from an AGB donor in NGC 188 than in M67, as the core masses are the same but the envelope masses are less in NGC 188 ($\sim$ 0.6 \Msun\ versus 0.8 \Msun, respectively). For the low-mass BSSs in NGC 188 to stand out as BSSs on the CMD, they still must accrete a similar amount of mass. Hence conservative and non-conservative mass transfer will transfer a more continuous distribution of mass, closing the gap. 

\cite{jadhavBlueStragglerStars2021} and \cite{carrasco-varelaMainSequenceBinary2025} examine BSS masses in comparison to their host cluster \Mmsto\ using the metric of fractional mass excess. Both of these papers use morphological definitions for the MSTO, effectively where an isochrone first reaches its bluest point after leaving the ZAMS. These definitions can lead to differences in \Mmsto\ compared to using the end of core hydrogen fusion, as we do in this work. Like \cite{leinerCensusBlueStragglers2021} and this work, both papers find that most BSSs are relatively low mass compared to \Mmsto\;(for \Mrel\;$<1.5$: \citealp[54\%,][]{jadhavBlueStragglerStars2021}; \citealp[81\%,][]{carrasco-varelaMainSequenceBinary2025}). \citeauthor{jadhavBlueStragglerStars2021} find this fraction to increase with cluster age, whereas \citeauthor{carrasco-varelaMainSequenceBinary2025} find no correlation with cluster age. \cite{jadhavBlueStragglerStars2021} found 7\% of BSSs in the oldest open clusters (log age $>$ 9.5) to have more than double the mass of \Mmsto. Neither \citeauthor{leinerCensusBlueStragglers2021} nor \citeauthor{carrasco-varelaMainSequenceBinary2025} identified any BSSs of such large mass ratio; we also do not.  

Both of these works also use the fractional mass excesses of BSSs to assign likely formation mechanisms, with BSSs of mass less than $1.5\rm{\;M_{MSTO}}$ being primarily formed by (non-conservative) mass transfer, BSSs of masses between $1.5 \rm{\;M_{MSTO}}\;and\; 2\rm{\;M_{MSTO}}$ being mostly formed by mergers or collisions, and more massive BSSs being formed through multiple mergers (by \cite{jadhavBlueStragglerStars2021}); neither paper examined BSSs of mass less than \Mmsto. They both acknowledge that fractional mass excess only suggests formation mechanism, with \citeauthor{carrasco-varelaMainSequenceBinary2025} noting that far-UV observations are needed to identify hot companions to distinguish different formation mechanisms. As shown in Figure~\ref{fig:bss_m67_n188}, at least 5 of 8 BSSs of \Mrel\;$>1.5$ in M67 and NGC 188 are securely due to mass transfer, including two that host WD companions. As such, we caution that additional information beyond mass is needed to reliably infer formation pathways. Further, although mergers or collisions may form BSSs of \Mrel\;$>1.5$, mass transfer still appears to be the dominant formation mechanism for these stars.  

\subsection{BSS Stellar Structure}\label{subsec:BSS_structure}

In Section~\ref{subsec:core}, we used single-star MESA models to determine the core helium content of stars in the same CMD positions as the BSSs in this work. Presuming the BSSs to have similar core helium content, we concluded that many BSSs have an appreciable build-up of helium in their cores at the time of their transformation, especially those of higher mass. Here, we discuss the few studies that have directly modeled the helium content in cores of low-to-intermediate mass-transfer products.

\cite{waggAsteroseismicImprintsMass2024} modeled a 3.0 \Msun\;star accreting 0.5 \Msun\;after it had already evolved on the main sequence. They found that after accretion, the star moved to a non-ZAMS location on the CMD and continued to evolve much the same as a single 3.5 \Msun\;star would, even though its core structure bore the marks of having initially been lower mass in the chemical gradient of its abundance profile. As the central temperature and convective core of the star grew during accretion to support the more massive and luminous star, it swept up hydrogen-rich material outside of the core, pushing the chemical gradient out to larger mass coordinates and increasing the core central hydrogen abundance, partially rejuvenating it (although it never returned to the primordial central hydrogen abundance). At the end of accretion, the star's central hydrogen fraction followed the same evolution as a single star of the same mass, but delayed by 72 Myr due to the initially slower stellar evolution of the lower mass star and the higher central hydrogen abundance post-accretion. 

M. Sun et al. (\citeyear{sunWOCS5379Detailed2021,sunSunWOCS4540Detailed2023}) created detailed models of two of the BSSs in Figure~\ref{fig:he_content}: WOCS 5379 through RGB mass transfer and WOCS 4540 through AGB mass transfer. The authors detail the stellar life-cycles of these stars based on best-fit models, finding that both moved to locations of (partially) evolved main-sequence stars having their new BSS masses. WOCS 5379 (\age = 0.26, mass = 1.21 \Msun) showed a small decrease in central helium abundance after accretion, but did not revert to primordial levels and then continued to evolve like a 1.21 \Msun\;star due to similar interior structures. The best-fit model of WOCS 4540 (\age = 0.91, mass = 1.5 \Msun) was a newly formed subgiant with a helium-rich core at the initiation of mass transfer. The central helium abundance was unchanged at the end of mass transfer. However hydrogen fusion continued in outer layers of the core, making the star an evolved main-sequence star before it eventually re-exhausted core hydrogen and continued on with stellar evolution as a single star of the same mass. These models indicate that mass transfer does not greatly change core helium abundance of the accretor star, and that using derived helium amounts in the cores of BSSs is, at least to first order, a reasonable metric of the evolutionary state of their accretor stars prior to formation. 

One potential issue of using single-star evolutionary tracks to model BSS evolution post-accretion is whether the final helium core mass of the BSS is the same as a single star of the same mass. In a study of accreting massive ($>11$ \Msun) main-sequence stars, \cite{schneiderPresupernovaEvolutionFinal2024} found that main-sequence stars that accreted $\leq 50\%$ of their initial mass (the same amounts the BSSs in the present study would have accreted) produced core helium abundances by the TAMS that were 0--15\% lower than single stars of the same mass as the BSS. This effect was more pronounced for stars that accreted little material or if the interaction occurred when the accretor was near the MSTO, due to steep chemical gradients that did not allow convective mixing of low-mean-molecular-weight material into the core. The BSSs studied here, while of much lower mass, are still primarily above the mass ($\approx1.1$ \Msun) at which convective cores develop, so they may show similar patterns to higher mass stars. Steep chemical gradients preventing convective mixing may provide an explanation for the small differences in helium-core content between the massive BSSs and their initial MSTO locations seen in Figure~\ref{fig:he_content}. If low-mass BSSs do not produce as much core helium as single stars of the same mass, evolution age may not accurately capture the remaining main-sequence lifetime of a BSS, as the same mass star would reach this lower core helium abundance earlier. This in turn would increase \age, shifting the population distribution even closer to the TAMS. 

Ultimately, wide-ranging detailed binary stellar evolution models are needed to establish the expected core helium amounts after an interaction in order to determine the state of the accretor. Further, asteroseismology observations of BSSs should reveal properties indicative of the cores of these systems before the episode of mass transfer \citep{waggAsteroseismicImprintsMass2024}, which could probe the predictions made here that the core helium amount of the accretor is responsible for the CMD location of the BSS at transformation. Some of the high mass BSSs in old clusters are in the $\delta$ Scuti or the $\gamma$ Doradus instability zones, whereas in clusters of $\sim$ 2 Gyr these zones contain some of the lower-mass BSS, enabling a test of these predictions for different mass accretors.

\subsection{Binary Evolution in NGC 188}\label{subsec:bss_progen}

\begin{figure*}[h!t]
    \centering
    \includegraphics[width=\linewidth]{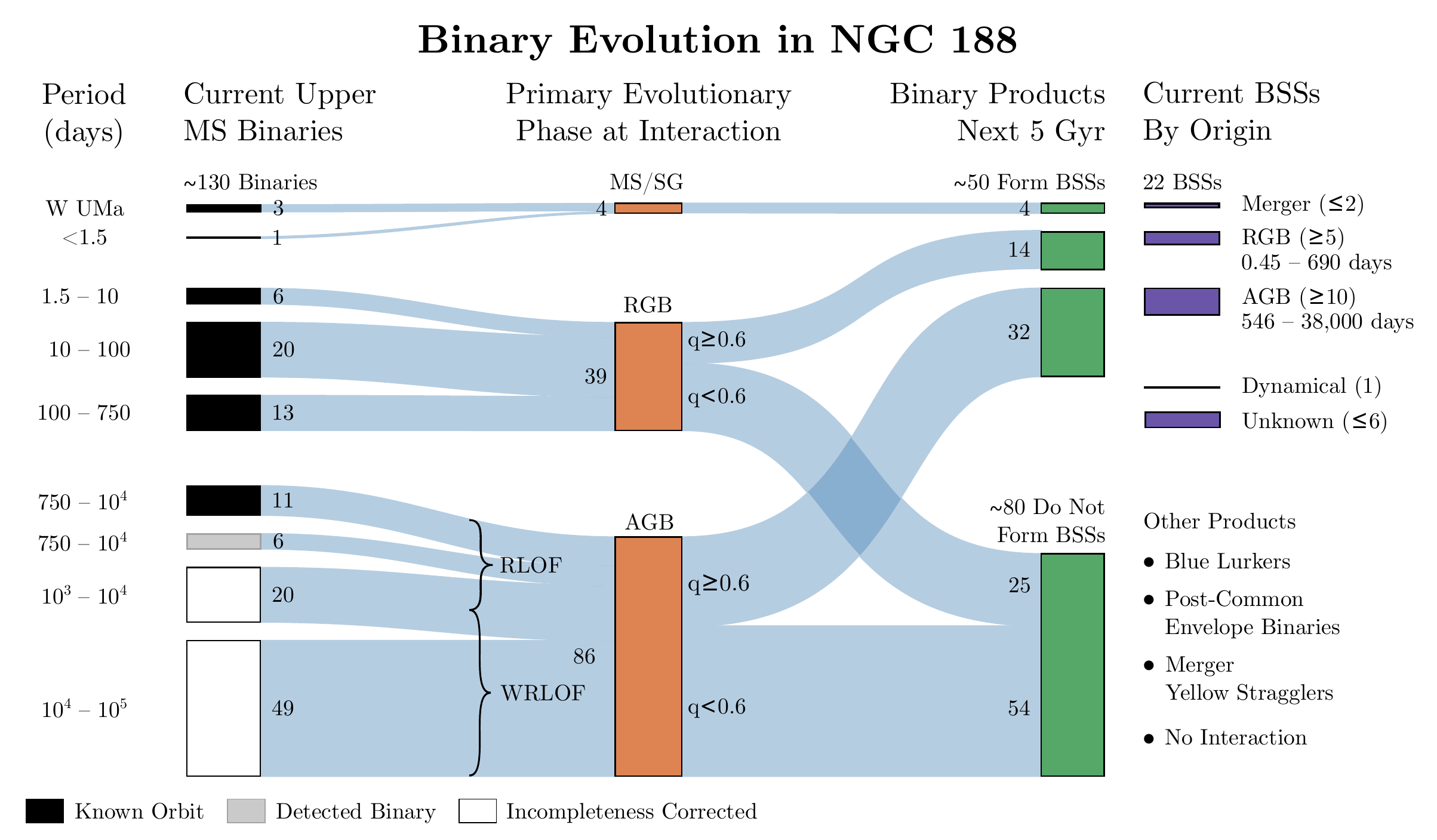}
    \caption{The evolution of the current binaries on the upper main sequence of NGC 188 into the next generations of BSSs. The current binaries of NGC 188 \citep{gellerWIYNOPENCLUSTER2009, narayanWIYNOpenCluster2026} are sorted by their orbital periods in the left column. Black boxes are binaries with known orbits, grey boxes those with radial-velocity variations indicative of having a period within that range, and the white boxes are incompleteness corrected according to \cite{moeMindYourPs2017}. The orbital periods of these binaries have a correspondence with the radius---and thus evolutionary phase---at which the primary would overflow its Roche lobe. We trace how binaries of different period ranges map to evolutionary phase in the second column (orange boxes). For AGB stars, we show approximate period ranges of binaries that could undergo RLOF or WRLOF \citep{sunWindRochelobeOverflow2024}. As found in Section~\ref{subsec:core}, RGB and AGB mass-transfer systems were generally created by progenitor binaries with $q>0.6$. The green boxes show how this limit propagates through to the number of binaries that will and will not form BSSs when they interact over the next 5 Gyr. Finally, we compare the predicted number of BSSs from among the current upper main-sequence binaries to the current generation of BSSs in the cluster (purple boxes, see Appendix~\ref{app:categorization} for details).}\label{fig:sankey}
\end{figure*}

At the age of NGC 188 (6.6 Gyr), the masses of stars near the turnoff and on the giant branch are not changing quickly over time. Thus the progenitor binary population of the current generation of BSSs---which all transformed within the last few Gyr---was essentially the same as the current binaries on the upper main sequence. Primary stars of these masses take about 1 Gyr from when they leave the main sequence until they reach the tip of the RGB and then another approximately 140 Myr to go through core helium burning and the AGB. Consequently, some of the current BSSs may have had progenitor donors with the same masses as giant stars in the cluster today. NGC 188 hosts one of the best-studied binary populations among old open clusters, with orbital parameters known for many upper main-sequence and brighter binaries, providing an invaluable case study for understanding how a binary population evolves into a BSS population.

\subsubsection{Current Binary Population}\label{subsubsec:binary_pop}

Using radial-velocity observations made over a baseline of $\sim10,000$ days, \cite{gellerWIYNOPENCLUSTER2009} and \cite{narayanWIYNOpenCluster2026} have derived orbital solutions for 16 of 22 BSSs and 51 of the 60 velocity-variable binary stars on the upper main sequence of NGC 188 (among the 266 upper main sequence stars of $G \leq 16$ mag, the radial-velocity survey completeness limit of \cite{narayanWIYNOpenCluster2026}; excludes photometric binaries with combined light of $G \leq 16$ mag but where the primary is fainter than this limit). Of the upper main-sequence binaries with spectroscopic orbital solutions, 1 has $P<1.5 $ days, 6 have $1.5\leq P<10$ days, 20 have $10\leq P <100$ days, 13 have $100\leq P<750$ days, and 11 have $P\geq750$ days. 6 of the binaries without orbital solutions have radial-velocity variations indicative of systems with periods more than several hundred days ($2\lesssim\sigma_{RV}\lesssim10\;\rm{km\;s^{-1}}$); the other 3 are W UMa stars, discussed next. 

The NGC 188 field includes many W UMa stars. Of the eight W UMa members examined in \cite{chenPHYSICALPARAMETERSTUDY2016}, three are not Gaia proper-motion members according to the analysis of \cite{narayanWIYNOpenCluster2026} and two are fainter than their completeness limit. Of the three W UMa members (WOCS 5337/ES Cep,\footnote{The radial velocity measurements of WOCS 5337 published in \cite{narayanWIYNOpenCluster2026} is of a bright component moving around the barycenter velocity of the system, whereas the W UMa components are too faint to reliably derive radial velocities.} WOCS 4989/V371, WOCS 5004/V370) on the upper main sequence, two have orbital solutions \citep{liuPhotometricInvestigationThree2011, zhuTHREECLOSEBINARIES2014} and the third (WOCS 5004) is a radial-velocity variable that suggests a short period. Adding these three systems to the above brings the total number of stars with $P<1.5$ days to 4.

\citeauthor{narayanWIYNOpenCluster2026} estimate that they have detected the binarity of $\sim75\%$ of all binaries in the magnitude range of $G \leq 16$ with $P < 10^4$ days. The incompleteness is almost entirely for binaries with $P\geq1000$ days \citep{gellerWIYNOPENCLUSTER2012}. Thus we estimate an additional 20 binaries in this period range. Even longer period binaries ($P > 10^4$ days) assuredly are also present among the undetected systems---as the hard-soft orbital period boundary for NGC 188 is greater than $10^7$ days \citep{gellerWIYNOPENCLUSTER2008,gellerINTERRUPTEDStelLARENCOUNTERS2015}---and may produce BSSs. For example,  \cite{sunWindRochelobeOverflow2024} found that wind Roche lobe overflow (WRLOF) can transfer material for stars in periods from 3000 days to 100,000 days. 

\cite{gellerWIYNOPENCLUSTER2012} found that the period distribution of NGC 188 main-sequence binaries follows the field period distribution of \cite{raghavanSURVEYSTELLARFAMILIES2010}. Given the observation of 10 binaries of $P<10^1$ days, 20 binaries with $10^1<P<10^2$ days, and $\gtrsim16$ binaries with $10^2<P<10^3$ days, the incompleteness and WD-companion corrections to the  field solar-mass period distribution of \cite{raghavanSURVEYSTELLARFAMILIES2010} by \cite{moeMindYourPs2017} yields $34\pm5$ binaries with $10^3<P<10^4$ days (consistent with our detection incompleteness correction above) and about $49 \pm 7$ binaries of $10^4<P<10^5$ days. 

Notably, extrapolating the number of short period systems ($P<10^3$ days) using the relative binary frequencies of each period decade \citep{moeMindYourPs2017} indicates that $48\pm3\%$ of the 266 stars on the upper main sequence are binary stars of $P <10^5$ days. The absolute frequencies found by \cite{moeMindYourPs2017} suggests that binaries of these periods should comprise $22 \pm 2\%$ of the total stellar population. The relative frequencies in each NGC 188 period decade match those of \cite{moeMindYourPs2017}. The higher overall frequency in NGC 188 may be a sign that binary stars have preferentially survived cluster evaporation compared to single stars \citep{gellerWIYNOPENCLUSTER2012}.

Although none of the main-sequence binaries in NGC 188 are known to have periods greater than a few thousand days, several of the very narrow-lined giants show radial-velocity variation over ten thousand or more days. The radial-velocity variation of these stars may be easier to identify in giants due to the very narrow lines of giants and because these stars would be much more luminous than a companion. 

\subsubsection{BSS and Binary Orbital Period Frequencies}

The evolutionary state of a progenitor donor at the advent of mass transfer is strongly linked to the binary orbital period when the donor fills its Roche lobe at periastron. Therefore, comparing the progenitor period distribution and the BSS period distribution can shed light on the formation pathways to BSSs. Solar-mass stars with periods less than about 1.5 days are expected to interact when the donor is on the main sequence or subgiant branch through mass transfer or through merger due to magnetic braking or the Kozai-Lidov mechanism driven by a tertiary star \citep[Figure 9 of][]{mathieuBlueStragglersFriends2025}. RGB mass transfer will occur out to periods at which a star at the tip of the RGB will overflow its Roche lobe \citep[$P< 750$ days in NGC 188,][]{narayanWIYNOpenCluster2026}. Finally, AGB mass transfer will occur at periods from 750 days to a hundred thousand days either through Roche lobe overflow (RLOF) or WRLOF. (\citealp[WRLOF becomes the dominant mechanism beyond periods of a few thousand days for circular orbits,][]{sunWindRochelobeOverflow2024}.) Tidal forces or winds on the giant branch can shrink the orbital periods of systems prior to interactions \citep{flemingRotationPeriodEvolution2019, ivanovaCommonEnvelopeEvolution2013} or a highly eccentric orbit can bring stars closer together at the same period than a circular orbit would, leading to interactions at an earlier evolutionary state then circular main-sequence orbital periods would indicate.

The upper main-sequence binaries in Section~\ref{subsubsec:binary_pop} have primary stars with masses of 1.0--1.1 \Msun; these will become giants over the next $\sim5$ Gyr. In Figure~\ref{fig:sankey}, we trace the evolution pathways that these progenitor binaries will take based on their current orbital periods. Most of the binaries ($67\pm7\%$) have orbital periods leading to interaction on the AGB, $30\pm5\%$ have orbital periods leading to interaction on the RGB, and only $3\pm2\%$ interact as subgiants or on the main sequence (including the currently interacting W UMa stars). 

Most of these binaries, however, will not produce BSSs. In Section~\ref{subsec:core}, we found that all but one of eleven RGB and AGB mass-transfer BSSs were created by progenitor binaries with $q>0.6$. \cite{childsStellarDynamicsOpen2025} find that the mass-ratio distribution of old open clusters, including NGC 188, follow the field mass-ratio distribution. Only about one-third of field solar-mass binaries with $0.2<\log P \,\rm{(days)}<5$ have companions with $q > 0.6$ \citep{moeMindYourPs2017}. Using this mass-ratio limit, we find that 14 RGB-mass-transfer and 32 AGB-mass-transfer BSSs will be created from the current binaries on the upper main sequence. 

The ratio of predicted RGB mass-transfer binaries to AGB mass-transfer binaries ($41 \pm 13\%$) is very similar to the ratio of RGB-origin to AGB-origin BSSs found in Section~\ref{subsec:bss_donors} ($37\pm18\%$). To achieve this agreement again requires that a large number of unobserved long-period progenitor binaries undergo mass transfer on the AGB. Specifically, WRLOF at very long periods must substantially contribute to BSS formation. The recently discovered eccentric 100-year orbit of the BSS WOCS 5020 in NGC 188 \citep{narayanWIYNOpenCluster2026} may provide an example of this.

Main-sequence mass transfer and mergers happen between close binaries of many mass ratios \citep{rasioMinimumMassRatio1995,stepienEvolutionCoolClose2011,stepienEvolutionLowMass2012}, so in Figure~\ref{fig:sankey} we show all 4 short-period progenitor binaries as creating BSSs. Among the current BSSs, only 2 can possibly be the result of mergers (WOCS 4306 and 4945, marked as unknown origin in Section~\ref{subsec:bss_donors}; see Appendix~\ref{app:categorization}). BSSs formed through collisions or exchanges through dynamical encounters are difficult to trace back to an original binary period, but are a small fraction of the total BSSs in old open clusters. We suspect that they are not a major contributor to the BSSs of unknown origin in Figure~\ref{fig:sankey}.  

Based on the progenitor binary population, the total number of BSSs that will be produced in NGC 188 over the next 5 Gyr is $\sim50$ (Figure~\ref{fig:sankey}). Mapping the number of current BSSs to the number of future BSSs is complex, as the population of BSSs is not replaced all at once, nor do all BSSs spend the same amount of time in the BSS region post-formation (e.g., see Section~\ref{subsec:core}). Further the production rates of different formation mechanisms are not well known. However, the current NGC 188 have total main-sequence lifetimes of 1--5 Gyr. With $\sim50$ BSSs predicted to be created in the next 5 Gyr, expecting $\sim25$ to be created in the next 2.5 Gyr is consistent with the 22 created in the last few Gyr. 

Finally, even though they are at periods that would interact, about 80 of the progenitor binaries will not produce BSSs due to their low mass ratios. At most, then, $39\pm5\%$ of the current upper-main-sequence binaries will go on to form BSSs. The other binaries may become blue lurkers or undergo common envelope evolution and then either merge into a yellow straggler or survive as a close binary containing a WD \citep{hjellmingResponseMainsequenceStars1991}. Whatever the outcome of each such interaction, the secondary star in the progenitor binary cannot gain enough mass to make it stand out in a CMD as a BSS. 

As discussed above, the total observed frequency of short-period binaries in NGC 188 is twice that found in the field. As a thought experiment, we consider the effect on the fraction of binaries becoming BSSs if the total binary frequency is 50\%, as found in the field \citep{moeMindYourPs2017}. Among the 266 stars on the upper main sequence 133 would be binaries, of which 60 have already been observed as $P<10^4$-day binaries. The binaries with $P<10^3$ days are very likely complete. We distribute the other conjectured 73 binaries among the $P>10^3$-day decades so the incompleteness-corrected numbers follow the relative frequencies of \cite{moeMindYourPs2017}. Of these, 42 binaries would interact on the AGB ($750<P<10^5$ days), half as many as in Figure~\ref{fig:sankey} and essentially the same number as the observed binaries with periods that will interact on the RGB. In order for these AGB binaries to form the observed AGB-origin BSSs, most of these would have to form BSSs when they interact, including those with $q$ as low as 0.3. Given the severe challenges this outcome would pose for mass-transfer stability, we suggest it far more likely that there actually are many unobserved binaries of $10^3<P<10^5$ days that do not produce BSSs when they interact.

In the story of how the binaries of NGC 188 give rise to its BSS population, two results emerge as particularly striking. First, only $\sim30-40\%$ of binaries that interact go on to form BSSs. This means that most other interactions do not transfer significant amounts of mass, so must be very non-conservative or unstable, perhaps going through common envelope evolution. Second, many AGB-origin BSSs are created by AGB stars in very long-period orbits. This indicates that either WRLOF from distant AGB stars and/or RLOF in very long period, highly eccentric orbits must contribute to the BSS population. Notably, although AGB mass transfer is the dominant formation mechanism for BSSs, current detection techniques are insensitive to the signals of such long-period binaries or require such long baselines of observation. Despite NGC 188 containing one of the best studied binary populations of any cluster, most of the actual progenitors of the next generation of AGB-origin BSSs cannot be identified.

The story seems largely the same in M67. Using the radial-velocity data and orbital solutions of \cite{gellerStellarRadialVelocities2021} and our 6-dimensional membership list, there are 561 stars on the upper main sequence within the completeness limit of G $<$ 15.5 mag, of which 186 have been identified as binary stars, including 84 with orbital solutions. \cite{gellerStellarRadialVelocities2021} estimate that 74\% of binaries with periods less than $10^4$ days have been detected and that 50\% of binaries with $P<3000$ days have known orbits, whereas no orbit solutions above 3000 days have been found. Following the categorization procedure of Section~\ref{subsubsec:binary_pop} for binaries with and without known orbits, $\sim2$ appear to be very short period, $\sim24$ have periods of several days, $\sim38$ have periods of tens of days, $\sim46$ have periods of hundreds of days, and $\sim77$ are greater than 1000 days, which is consistent with the orbital period distribution of NGC 188, particularly at long periods. As found in Section~\ref{subsec:bss_donors}, AGB mass transfer is the dominant formation mechanism in this cluster, so again many of the binaries that would undergo AGB RLOF or WRLOF have probably not been identified.

Examining the stars that will ascend the giant branch over the next 2 Gyr (essentially until the age of NGC 188 with \Mmsto\ of 1.1 \Msun), 68 of 223 stars have been identified as binaries. These binaries have orbital periods that will lead to an interaction, with likely more at very long periods that would undergo WRLOF. The BSSs created by these binaries will have masses that will live for several billion years as BSSs, and so they represent a complete progenitor binary population for the generation of BSSs when M67 is the age of NGC 188. Although M67 is about 40\% more massive than NGC 188 \citep{alvarez-baenaLongevityOldestOpen2024}, and thus may have more stars that could produce BSSs, the number of binaries that will interact in the next 2 Gyr is several times larger than the number of BSSs that currently exist in either M67 (19) or NGC 188 (22). This once again indicates that most binaries will not produce BSSs after interacting.

\subsubsection{BSS and Binary Orbital Period Evolution}

As seen in Figure~\ref{fig:bss_m67_n188}, the frequency of BSSs drops dramatically below periods of approximately 100 days, even though radial-velocity surveys are quite sensitive to binaries with periods in this range \citep[see also Figure 4 and discussion in Section 3.4.2 of][]{mathieuBlueStragglersFriends2025}. Specifically in NGC 188, no BSSs have periods between 10 and 100 days. Yet $\sim16\pm3\%$ of the upper main-sequence binaries in NGC 188 that will interact are in this period range. The short-period distribution of main-sequence binaries and the BSSs of NGC 188 (and M67) are very different---a property first noticed by \cite{mathieuBinaryStarFraction2009}. Here we examine what the orbital periods of BSSs can tell us about the period distribution of their progenitors and how it changed during interactions. 

Main-sequence binaries in the orbital period range of 10--100 days will interact on the RGB. By combining MIST stellar radii for giant stars in NGC 188, Kepler's third law, and the Eggleton approximation of the Roche lobe radius \citep[Equation 2 of][]{eggletonAproximationsRadiiRoche1983}, we find that equal-mass binaries in circular orbits with helium core masses between 0.2 and 0.35 \Msun\ would overflow their Roche lobes at periods between 5 and 100 days. Combining our findings in Section~\ref{subsec:bss_donors} that all but one of the RGB-origin BSSs has a statistical secondary mass within this range ($0.2<m_2\sin(60^\circ)<0.35$) and in Section~\ref{subsec:core} that most mass-transfer systems were created by progenitor binaries with $q>0.6$---which corresponds to $0-0.4$ \Msun\ below \Mmsto\;in these clusters---implies that the progenitor binaries of the NGC 188 RGB-origin BSSs had periods before the interaction precisely in the observed BSS period gap (barring effects of tides or mass-loss through winds).

A possible explanation for the BSS period gap is that in these old clusters RGB (and AGB) stars are only about 0.05 \Msun\ more massive than \Mmsto. In binaries with $q>0.6$, progenitor donors only need to lose 0.05--0.4 \Msun\ before the binary mass ratio flips. At this point, conservation of angular momentum requires the binary to widen and the orbital period to increase, which creates a stabilizing effect on mass transfer. (Non-conservative mass transfer complicates this based on whether mass is lost from the vicinity of the accretor or donor, but will also widen orbits.) Moreover, the donor giant still has more envelope to lose, which extends the binary to even wider orbits. The result is to turn progenitor binaries with periods of less than 100 days into BSSs with periods longer than 100 days, forming the observed period gap among the BSSs. 

\cite{petrovicInfluenceInitialOrbital2021} performed MESA simulations of low-mass-binary ($m_1=1.2$, $m_2=1.0$ \Msun) mass transfer at various efficiencies $\beta$ for periods from 10--100 days. They found that binaries with periods of 50--100 days underwent stable mass transfer at all $\beta$ and had final orbital periods of 300--500 days. Only the most non-conservative mass transfer ($\beta\leq0.2$) was able to produce non-contact BSSs with orbital periods at the start of mass transfer of 10 days. After mass transfer these systems had final periods of just over 100 days. Although examining high-mass binaries (7--120 \Msun), \cite{rochaMassTransferEccentric2025} also found that equal-mass systems with initial periods between 10 and 100 days would grow to longer periods by the conclusion of mass transfer. 

Two of the RGB-origin BSSs with $\sim0.2$ \Msun\ companions in Figure~\ref{fig:bss_m67_n188} (WOCS 1007 in M67 (\age=0.78, \Mrel=1.58) and WOCS 4230 in NGC 188 (\age=0.69, \Mrel= 1.08)) have periods of less than 10 days, compelling evidence of having undergone a period of unstable mass transfer and gone through common envelope evolution. The accretors in both cases were nearing or at the MSTO, which could indicate that the reason these systems survived without merging was that $q$ was high enough that common envelope ejection did not remove enough angular momentum from the orbit to drive the stars together. In Section~\ref{subsec:core}, we found that for BSSs with known WD companions, low-mass BSSs had gained $0.15-0.4$ \Msun~and high-mass BSSs gained $0.5-0.8$ \Msun. RGB-origin BSSs were among both groups with with WOCS 4230 in the former and WOCS 1007 in the latter, indicating that they may have had stable mass transfer of different $\beta$ of  prior to going through a common envelope phase. 

Next, we consider the $10 \pm 3\%$ of main-sequence binaries that will interact and have orbits between 100--750 days. There is only one RGB-origin BSSs (of 5 in NGC 188 and the 7 in this study) that has a secondary mass indicative of mass transfer beginning at a period above 100 days. In the previous sub-subsection, we found that most stars that would interact on the RGB cannot produce BSSs due to the numbers of binaries with relevant periods versus BSSs. These stars will interact but must go through common envelope evolution as the secondary does not accrete sufficient mass to stand out as a BSS \citep{hjellmingResponseMainsequenceStars1991}. We conjecture that the low minimum helium WD masses of the BSSs is evidence that the stars of 100--750 day periods are more likely to go through common envelope evolution than those of 10--100 days. This may be supported by theory; if RGB envelopes have a superadiabatic response to mass transfer, mass transfer can be stable for giant donors with small cores \citep{ivanovaBinaryEvolutionRoche2015}.  

As noted, the one BSS from this period range, WOCS 5379 ($P=120$ days, \age = 0.26, \Mrel = 1.08), has a helium WD companion with a measured mass of 0.42 \Msun\ \citep{gosnellConstrainingMasstransferHistories2019}. For an equal-mass 1.15 \Msun\ binary in a circular orbit, this core mass would correspond to a 350-day period, three times longer than its current period. \cite{sunWOCS5379Detailed2021} modeled this system with stable mass transfer in MESA and found their best-fit model to start at $P = 12.7$ days and end at $P=120$ days, but it underestimated the donor's core mass (their models produced a WD of 0.33 \Msun), potentially indicating that system may have also undergone a period of unstable mass transfer.

\cite{temminkCopingLossStability2023} examined several mass-transfer stability criteria and found mass transfer generally became unstable when the donor's response to mass loss became quasi-adiabatic. Using MIST and Figure 8 of \cite{temminkCopingLossStability2023} for a 1.1 \Msun\ RGB star, periods that correspond to RLOF during the later portion of the subgiant branch and the beginning of the RGB were generally unstable at $q\leq1$. However, at periods of 10--100 days ($0.8 < \log (R/R_\odot) < 1.6$), the critical mass ratio fell from $q = 1.0$ to $q=0.6$, indicating that only binaries of $q>0.6$ would undergo stable mass transfer at such periods. At larger donor radii extending to the tip of the RGB, they found that donors may overflow an outer lobe of their Roche potential or undergo dynamical-timescale evolution if $q<0.6$, which could lead to unstable mass transfer and shrink orbits. These stability boundaries are consistent with our observational findings above.

Altogether, the RGB-origin BSSs of NGC 188 (and M67) tell a variety of stories. Unlike very long-period systems where the exact boundaries of interaction are still unclear, all of the binaries with periods less than 750 days ($33\pm5\%$ of the total predicted binary population of $P<10^5$ days) will overflow their Roche lobes and interact; most will not form BSSs during this interaction, indicating those binaries likely undergo common envelope evolution. Based on statistical secondary masses, all presumed to be WDs, most BSSs underwent RLOF as they went up the RGB rather than near the tip of the RGB. These stars all gained mass to become BSSs, establishing that they underwent stable mass transfer. However, 3 of 7 have periods that also indicate they underwent unstable mass transfer as well. Together, these suggest that the majority of RGB binaries at these periods go through a common envelope phase in their evolution.

AGB stars too must undergo orbital evolution to match our analysis of the incompleteness-corrected orbital period distribution in the previous sub-subsection. In order to produce the number of AGB-origin BSSs in NGC 188 ($\geq10$), there have to be as-yet-unobserved very-long-period binaries that will make BSSs, either through RLOF in very eccentric systems or through WRLOF. Yet, in the current population of BSSs in NGC 188, 8 have orbits in the many hundred days to few thousand days range, with only up to an additional 4 of 22 that could have $P>10^4$ days (either from the known orbit or from low-or-no radial-velocity variability, although one could also be a single-star merger BSS). This indicates that very long-period systems must shrink their orbital period during the interaction.

\cite{rochaMassTransferEccentric2025} has shown that RLOF mass transfer can occur among these very long period binaries provided they have high enough eccentricity. Depending on the mass ratio and initial eccentricity, the final period of these binaries can shrink to several thousands of days or even grow to many tens of thousands of days. Orbital eccentricity too can change dramatically, including being pumped up once the mass ratio is near unity. This may further slow mass transfer, leading to more stable mass transfer in these systems. 

In these old open clusters, progenitor donors have degenerate cores that take tens of millions of years to significantly change mass, which means their WD masses tie them to a specific evolutionary state at the point of mass transfer. This also ties them to specific orbital periods shortly prior to the onset of mass transfer by examining when the progenitor giant radius would overflow its Roche lobe. Building on this, we found evidence that the orbits of the BSSs may be a dex different than the progenitor binaries that created them. Among the BSSs in NGC 188 and M67 that interacted on the RGB, those with initial orbits of 10--100 days (from WD minimum masses) yielded BSSs with final periods in the range of 100--700 days if they underwent just stable mass transfer or less than 10 days due to undergoing common envelope evolution. Those with initial orbits of 100--1000 days shrank due to an episode of common envelope evolution. Common envelope evolution may be very common among this latter population as the progenitor binaries of these periods are as prevalent in the cluster as those of the former group, but only 1 of 7 RGB-origin BSSs in this study has a minimum mass indicative of having an initial period greater than 100 days. Among the AGB-origin BSSs, very long-period binaries must contribute to the population in order for AGB mass transfer to be the dominant mechanism of BSS creation; yet, most of the AGB-origin BSSs have periods around 1000 days. As discussed by \cite{gellerWIYNOPENCLUSTER2012}, the NGC 188 main-sequence binary population periods match the period distributions of galactic field binaries, indicating that the results discussed here about the progenitors of the NGC 188 BSSs also apply to field BSSs of similar mass. 

\subsection{Implications for Observational Surveys of BSSs}\label{subsec:obs_implications}

In this work, we showed that most massive BSSs in old clusters are in the final third of their main-sequence lifetimes and found evidence that this is because the progenitor accretors of these stars were near the end of their main-sequence lifetimes at the time of formation. In these old clusters, massive BSSs are approximately $1.5-2.2$ \Msun (A- and F-type), which, as single stars, would have total main-sequence lifetimes of $\sim1-2$ Gyr. This implies that we should only expect to observe them as BSSs for several hundred Myr after their transformation. In younger clusters, BSS masses can exceed 3 \Msun and have main-sequence lifetimes of only a few hundred Myr, suggesting that such stars may remain observable as BSSs for $\sim100$ Myr, or even less for more massive stars. Further, equal-mass binaries, which show a slight population excess \citep{moeMindYourPs2017}, could create the most massive mass-transfer BSSs, which would have the shortest observable time frames due to their very enriched cores and high masses. In turn, this indicates that population studies, including this one, could be biased toward lower-mass BSSs: not only do these stars have longer main-sequence lifetimes than higher-mass BSSs, but they can also form closer to the ZAMS, so remain BSSs for significantly longer periods. As a consequence, formation mechanisms that preferentially produce the most massive BSSs may be systematically under-counted, particularly among those that require a massive accretor. Additionally, the most massive stars in a cluster may be the most difficult to identify a WD companion next to due to their luminosities and UV flux, further complicating understanding how these stars formed.

Further, this work relied on long-baseline radial-velocity surveys to identify the formation mechanisms of many BSSs. It is possible that the current generation of radial-velocity surveys (errors on order of $0.5\rm\; km\;s^{-1}$ for slowly rotating stars) are not sensitive to low inclination, thousand-plus-day orbits of CO WDs produced by AGB mass transfer. In the top panel of Figure~\ref{fig:bss_m67_n188}, we note that no BSSs have been found in these clusters with periods above 1000 days and with minimum secondary masses below 0.4 \Msun. For a nominal BSS mass of 1.5 \Msun, CO WD mass of 0.55 \Msun, 1800-day, $e=0.1$ orbit, and inclination of $20^\circ$ ($m_2\sin i \sim0.2\rm \;M_\odot$), the system would have a semi-amplitude velocity of $\sim 2\rm\; km\;s^{-1}$, a signal that would only rise above detection limits if regularly observed across an entire orbital period of 5.5 years. Despite the $\sim2000$-day baseline, the upcoming Gaia data release 4 of time-series radial velocities may also not be sensitive to such systems, as the Gaia DR3 categorization of the Gaia RV spectrometer \citep{katzGaiaDataRelease2023} suggests that stars of $6000 < T_{\rm Eff} < 9000$ K---roughly the temperature range of most BSSs---will have precisions (much) higher than 1 $\rm km\;s^{-1}$ at the apparent magnitudes of the BSSs of the clusters in this study. 

\section{Summary}\label{sec:summary}

In this work we explore what the CMD distributions of BSS populations in old open clusters reveal about BSS progenitors and formation histories. BSSs in old open clusters are a superb population with which to study binary evolution pathways: determination of properties such as age, metallicity, reddening, and distance are informed by the host cluster; open clusters are low density and enable long-baseline time-series spectroscopic surveys of many stars (both BSSs and other cluster binaries); stars have generally spun down by these ages, enabling measurement of precise radial velocities and abundances; and stars of these ages have unique properties that enable probing important physics (e.g., degenerate core masses enable determination of donor evolutionary states and mass-transfer efficiencies). We use single-star MIST models to characterize donor and accretor stars in binaries and to investigate BSS interiors and evolution.

We find that half of the BSSs are at locations in the CMD corresponding to single stars in the last third of their main-sequence lifetimes. Few BSSs are on the ZAMS. This effect is primarily, but not solely, driven by the most massive BSSs in each cluster. The overpopulation of BSSs toward the TAMS indicates that stars enter this population already partly evolved.

Eleven of the BSSs in this study have WD companions with measured cooling ages, enabling us to determine their transformation ages. Very few of these BSSs formed on the ZAMS, but rather formed at all evolution ages. Using MESA models, we find that the amount of helium fused by a main-sequence star by the time it has reached the MSTO is equivalent to the amount of helium that a single star as massive as the  BSSs creates at approximately two-thirds of the way through its main-sequence lifetime. In other words, current BSSs have core helium amounts that can only be produced since their formation if their progenitor accretors already had helium amounts in their cores well in excess of primordial. Thus, many BSSs in old open clusters are created from MSTO stars and start their evolution far along their associated main-sequence evolutionary tracks, explaining the build-up of BSSs toward the TAMS. 

We further show that AGB mass transfer creates massive BSSs through conservative mass transfer onto these helium-enriched MSTO stars. Low-mass BSSs (and possibly blue lurkers) can be formed by non-conservative mass transfer from AGB stars onto lower-mass main-sequence stars, and thus form over a wider range of evolution ages. BSSs formed from the lowest-mass main-sequence stars will begin their BSS evolution on the ZAMS, as such accretors have little helium enrichment.

Additionally, we find that the orbital periods, minimum secondary masses, and abundances of BSSs demonstrate that mass transfer from AGB and RGB stars is the dominant formation mechanism in old open clusters (at least two-thirds of BSSs) and that AGB mass transfer accounts for at least half of BSSs. We further find that RGB-origin BSSs follow roughly the same distribution in \age\;and \Mrel\;as AGB-origin BSSs, which suggests that they undergo the same formation processes as AGB stars.

We find that the period gap in BSSs between 10 and 100 days can be explained by binaries with those periods migrating to longer orbits during stable mass transfer or to shorter orbits during unstable mass transfer. AGB-origin BSS must also undergo significant orbital period migration to account for the current BSS orbits.

For the very well-studied cluster NGC 188, we compare the upper main-sequence binary and BSS orbital-period distributions. We find that in order for AGB-origin BSSs to be the majority of the BSS population, there must be many unobserved long-period binaries that would undergo AGB WRLOF or (highly eccentric) RLOF. Crucially, at most 30-40\% of binaries with $P<10^5$ days will form BSSs when they interact.

In summary, we show that the CMD distributions of BSSs in old open clusters, when combined with orbital solutions, abundance measurements, and analyses of WD companions, provide strong constraints on the progenitors and formation pathways of BSSs. We find that all mechanisms that are predicted to create BSSs---i.e., mass transfer (both conservative and non-conservative), mergers, and collisions---must happen to explain distributions of BSS orbital periods and companions, but that most form through mass transfer and that AGB mass transfer is the dominant mechanism. In many cases the progenitor accretors had to be enriched in core helium prior to interaction, pointing to progenitor binaries with companions near the MSTO. This in turn points to many progenitor binaries with roughly equal-mass stars, which are thought to be the most stable to RLOF mass transfer. 

\begin{acknowledgements}
The authors express their gratitude to D. Dixon, N. Leigh, E. Leiner, E. Motherway, R. S. Narayan, A. Nine, A. Quitral-Pierart, R. Townsend, and T. von Hippel for their insightful feedback during the creation of this manuscript and to the many undergraduate and graduate students of the R. D. Mathieu research group and the staff of WIYN observatory, without whom we would not have been able to collect thousands of stellar spectra that enabled the findings of this work. Finally, we acknowledge the support of the Wisconsin Alumni Research Fund and the Wisconsin Space Grant Consortium through awards RFP25\_4-0 and RFP25\_12-0.

This work has made use of data from the European Space Agency (ESA) mission Gaia (\url{https://www.cosmos.esa.int/gaia}), processed by the Gaia Data Processing and Analysis Consortium (DPAC; \url{https://www.cosmos.esa.int/web/gaia/dpac/ consortium}). Funding for the DPAC has been provided by national institutions, in particular the institutions participating in the Gaia Multilateral Agreement. 

This work was conducted at the University of Wisconsin-Madison, which is located on occupied ancestral land of the Ho-Chunk people, a place their nation has called Teejop since time immemorial. In an 1832 treaty, the Ho-Chunk were forced to cede this territory. The university was founded on and funded through this seized land; this legacy enabled the science presented here. Observations for this work were conducted at the WIYN telescope on Kitt Peak, which is part of the lands of the Tohono O’odham Nation.
\end{acknowledgements}

\facilities{WIYN - Wisconsin-Indiana-Yale-NOAO Telescope (Hydra MOS), Gaia}

\software{\textsf{Astropy} \citep{astropycollaborationAstropyCommunityPython2013,astropycollaborationAstropyProjectBuilding2018,astropycollaborationAstropyProjectSustaining2022}, \textsf{MIST} \citep{dotterMESAISOCHRONESLAR2016,choiMESAISOCHRONESSTELLAR2016,paxtonMODULESEXPERIMENTSSTELLAR2011,paxtonMODULESEXPERIMENTSSTELLAR2013,paxtonMODULESEXPERIMENTSSTELLAR2015}, \textsf{isochrones} \citep{mortonIsochronesStellarModel2015}, \textsf{NumPy} \citep{harrisArrayProgrammingNumPy2020},  \textsf{SciPy} \citep{virtanenSciPyFundamentalAlgorithms2020}, \textsf{scikit-learn} \citep{scikit-learn}, \textsf{emcee} \citep{foreman-mackeyEmceeMCMCHammer2013},  \textsf{dustmaps} \citep{greenGalacticReddening3D2018}, \textsf{ChatGPT-4} \citep{openai_chatgpt_2025}.} 

\appendix
\section{Discussion of Individual BSSs in M67 and NGC 188}\label{app:categorization}
\subsection{BSS Categorizations}
In the context of BSS formation by stable mass transfer, final orbital period can be diagnostic of the evolutionary state of the donor star, with AGB mass-transfer products generally having orbital periods longer than 1000 days (although some may be as short as 300 days) and RGB mass-transfer products generally having orbits between 100 and several hundred days (although potentially up to 1000 days) \citep{mathieuBlueStragglersFriends2025}. BSSs that have undergone an episode of unstable mass transfer after a period of stable mass transfer may have much shorter periods. In old open clusters another diagnostic can be added: secondary minimum mass, under the presumption that the companion is the WD remnant of the donor core at the time of BSS formation. Specifically, the progenitor donor masses of all BSSs in M67 and NGC 188 are less than 2 \Msun, meaning that the cores of the donor RGB and AGB stars were degenerate. From the MIST evolutionary tracks of solar-metallicity stars, the helium flash occurs when the core of an RGB star reaches 0.47 \Msun. 

We calculate the secondary minimum mass for each BSS system from the binary mass function and BSS mass. As this is a minimum mass, any secondary with mass $> 0.47\;$\Msun\;must be a CO WD from an AGB star progenitor. As the inclination angles ($i$) of most of these systems are unknown, we use the expected median $i = 60^\circ$ from a uniform distribution of $\cos(i)$ to classify BSSs with $m_2 > 0.47/\sin(60^\circ) = 0.41$ \Msun\;as having a CO WD companion. Using orbital period and statistical secondary mass alone, we identify 5 BSSs of RGB origin (M67: WOCS 3001; NGC 188: WOCS 4589, 5350, 5379, 5467) and 13 BSSs of AGB origin (M67: WOCS 1025, 3010, 3013, 9005; NGC 188: WOCS 2679, 4348, 4540, 4581, 4970, 5020, 5325, 5434, 8104). The secondary minimum masses reported here of WOCS 1025 (S1195) and 3013 (S752) are slightly higher than the mass of a CO WD of relevant AGB masses in M67, but small errors on semi-amplitude velocity (unestimated in original study) and the errors on BSS mass from this study produce secondary minimum masses that overlap the expected CO WD mass range, so we classify them as of AGB origin. (Additionally, WOCS 3013 is barium enriched, as discussed below, further suggesting it to be of AGB origin). 

We classify an additional 7 BSSs as resulting from AGB mass transfer (M67: WOCS 1020, 2011, 2013, 3005, 6038, 11006; NGC 188: 4290). Four of these stars show steadily decreasing or increasing radial-velocity trends $>$~1 km s$^{-1}$ over 7000 days using WIYN high-precision measurements \citep{gellerStellarRadialVelocities2021,narayanWIYNOpenCluster2026}, indicative of very long period orbits (M67: WOCS 1020, 6038, 11006; NGC 188: 4290 (with WOCS 11006 having a radial-velocity variation suggestive of a highly eccentric orbit). BSS abundance measurements of s-process element excesses are secure diagnostics of mass transfer from AGB stars of $M>1.3\;\rm M_\odot$ ~\citep{nineWIYNOpenCluster2024}.
\cite{nineWIYNOpenCluster2024} and
\cite{palDiscoveryBariumBlue2024} measured Ba excesses in five of the M67 BSSs. As expected, three of these (WOCS 2013, 3013, 9005) have long orbital periods. Two BSSs (WOCS 2011 and 3005) have not been detected as radial-velocity variables, but could be very long-period systems. Abundance measurement has not proven a useful diagnostic for other formation channels.

Below we discuss the formation mechanisms and categorizations of 7 additional BSSs based on detailed studies of these stars from other works: 

\textit{WOCS 1007} (S1284) in M67 is in a 4.2-day eccentric orbit and was found to be 2.03~\Msun\ in the CMD analysis here. \cite{vernekarPhotometricVariabilityBlue2023} found it to have a hot WD companion of 0.22 \Msun. They note that this companion mass corresponds to the core mass of an early RGB star, indicating it underwent mass transfer from an RGB star. We thus consider this star to be an RGB mass-transfer product. The high mass of the BSS in comparison to \Mmsto~but short orbital period suggests it underwent a period of stable mass transfer before a common envelope formed. 

\textit{WOCS 2009} (S1082) in M67 contains two BSSs in a hierarchical quadruple system, with one in a 1.07-day orbit binary and the other in a 1189-day orbit binary. \cite{quitral-pierartMassluminosityAnomaliesPlausible2025} argue that the system is of dynamical origin with the BSSs exchanged in, so the origins of the two BSSs are unclear. We do not include it in any figure, but do count it toward the population statistics in this section.

\textit{WOCS 4003} (S1036) in M67 was identified as a W UMa binary with a photometric period of 0.44 days by \cite{gillilandTimeResolvedCCDPhotometry1991}. As such, we categorize it here as main-sequence mass transfer. \cite{vernekarPhotometricVariabilityBlue2023} subsequently found an additional eclipse signal at 1.11 days, and suggested the system to be a compact triple, which could be driving the inner binary to interact, but also may be semi-detached. Re-analyzing the cross-correlation functions of \cite{gellerStellarRadialVelocities2021}, we derived a single-lined circular orbital solution with a period of $0.441432 \pm 0.000002$ days (Appendix~\ref{app:4003}). The residuals to this solution (often $>10$ km s$^{-1}$) are far larger than our measurement uncertainties (mostly $<5$ km s$^{-1}$), possibly indicative of a close companion. However, we do not find any evidence of periodicity in the residuals. If this system has ongoing mass transfer, its CMD position may not be representative of its evolution age. Because this star is 1 of 184 BSSs under consideration in Section~\ref{sec:mass_ages} and 1 of 41 BSSs in M67 and NGC 188, removing it from the sample does not meaningfully change results.

\textit{WOCS 4230} in NGC 188 is a very short period binary system (P=0.456 days) with a circular orbit \citep{narayanWIYNOpenCluster2026} and hot WD companion \citep{gosnellIMPLICATIONSFORMATIONBLUE2015}. \cite{narayanWIYNOpenCluster2026} found the companion to have a dynamical mass of 0.3 \Msun\ and noted that this indicates it to be of RGB origin. Like with WOCS 1007, it is possible that this system underwent stable mass transfer before a common envelope formed.

\textit{WOCS 4306} in NGC 188 is a non-velocity-variable BSSs and the best merger candidate in the cluster. It is also possible that this star could have a very distant companion that is causing radial-velocity variation below our threshold of detection (roughly, semi-amplitude velocities $ = 2\;\rm{km\;s^{-1}}$). As such we label it as of unknown origin in Section~\ref{subsec:bss_donors} but discuss it as merger candidate in Section~\ref{subsec:bss_progen}.

\textit{WOCS 4945} in NGC 188 is a faint BSS with of P = 5510.0 days and e = 0.48 \citep{narayanWIYNOpenCluster2026}. As discussed by \cite{narayanWIYNOpenCluster2026}, it has a minimum secondary mass of 0.75 \Msun. This minimum mass is significantly larger than the core mass of giants in NGC 188 (corresponding to the core mass of a 3 \Msun\ star). If the companion is not a WD, then a merger is the likely formation path of the BSS. If the companion is a WD, then it could have been the result of a merger that created a 3 \Msun\ BSS that then underwent AGB mass transfer on to the present-day BSS given the long orbit. Because we are unable to further constrain the formation mechanism of this star, we categorize it as unknown origin here.

\textit{WOCS 5078} in NGC 188 is a double-lined spectroscopic binary in a 4.7 day eccentric orbit. The binary comprises a BSS and a main-sequence companion. \cite{mathieuBinaryStarFraction2009} found this system to comprise a 1.5 and 1.02 \Msun\ BSS-main-sequence binary from dynamical masses and proposed this system to have a dynamical origin, with the BSS being created by an unknown mechanism before exchanging its companion. They note that the BSS mass is higher than the mass derived from photometry of 1.3 \Msun (we found it to have a mass of 1.34 \Msun). Extended radial-velocity monitoring of this system in \cite{narayanWIYNOpenCluster2026} does not show any long-term trends in RV that would be indicative of a nearby tertiary member. 

Finally, we were unable to classify 10 BSSs based on available information (M67: WOCS 1006, 1026, 4006, 7026, 8006; NGC 188: WOCS 4306, 4447, 4535, 5885, 5934). They include both velocity-variable (M67: WOCS 1006, 1026, 4006; NGC 188: WOCS 4447, 4535, 5885, 5934) and non-velocity-variable stars (M67: WOCS 7026, 8006; NGC 188: WOCS 4306), rapid rotators and narrow-lined stars, and two stars with hot companions (WOCS 4006 in M67 has a WD, WOCS 5885 may have a post-AGB companion \citep[not included in Figure~\ref{fig:bss_m67_n188}][]{subramaniamHotCompanionBlue2016}). None of these BSSs have known orbital periods. They likely include merger or collision products, and may also include additional long-period mass-transfer systems with velocity variability masked by rapid rotation. None of the velocity-variable BSSs have radial-velocity ranges suggesting periods on order of several days or less (roughly, semi-amplitude velocities $\rm>30\;km\;s^{-1}$), so these BSSs do not have main-sequence mass-transfer origins. However, it is possible that some of the merger BSSs went through a prior phase of main-sequence mass transfer. 

\subsection{M67 WOCS 4003 Orbit}\label{app:4003}

\begin{figure}
    \centering
    \includegraphics[width=0.5\linewidth]{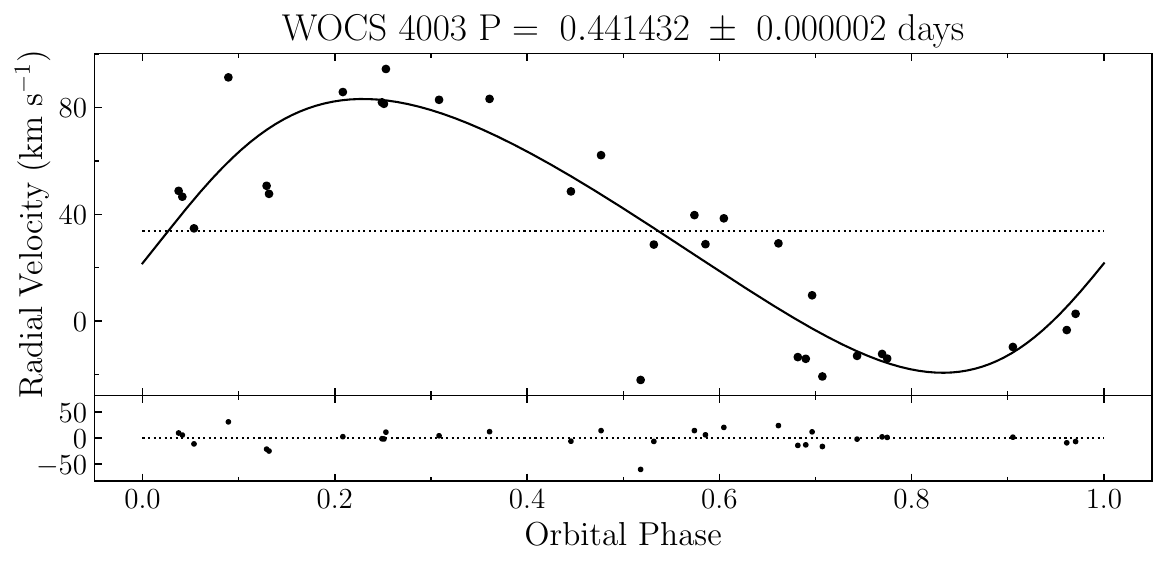}
    \caption{We plot radial velocity against orbital phase for WOCS 4003, showing the data points with black dots and the orbital fit to the data with the solid line; the dotted line marks the $\gamma$-velocity. Beneath each orbit plot, we show the residuals from the fit.}
    \label{fig:4003}
\end{figure}

The orbital solution for WOCS 4003 is given in Table~\ref{table:4003_orbit} and shown in Figure~\ref{fig:4003}. If this is a triple system, long-term $\gamma$ variations due to a triple member can appear as an increase in eccentricity. Table~\ref{table:4003_rv} gives the WOCS ID, Heliocentric Julian Date (HJD), the measured radial velocity, the residuals ($O - C$), orbital phase, and the cross-correlation function height (with a maximum of 1) of every spectroscopic measurement made of WOCS 4003 and WOCS 6038 by \cite{gellerStellarRadialVelocities2021}, although none were published in that work. These data were collected on the Hydra Multi-Object Spectrograph on the WIYN 3.5m telescope at Kitt Peak, Arizona.\footnote{The WIYN 3.5m Observatory is a joint facility of the University of Wisconsin–Madison, Indiana University, NSF’s NOIRLab, the Pennsylvania State University, and Princeton University.} Because WOCS 4003 is rapidly rotating (projected rotational velocity of $\sim105\;\rm{km\;s^{-1}}$), errors on individual radial-velocity measurements are on order of $5\;\rm{km\;s^{-1}}$; WOCS 6038 has errors of $0.4\;\rm{km\;s^{-1}}$.  

\begin{deluxetable*}{ccccccccccc} \label{table:4003_orbit}
\tablewidth{0pt}
\tabletypesize{\scriptsize}
\tablecaption{Orbital Parameters of M67 WOCS 4003}
\tablehead{
   \colhead{P} & \colhead{Orbital} & \colhead{$\gamma$} & \colhead{K} & \colhead{$ecc$} & \colhead{$\omega$} & \colhead{T$_0$} & \colhead{$a \; \sin(i)$} & \colhead{$f(m)$} & \colhead{$\sigma$} & \colhead{N} \\
    \colhead{days} & \colhead{Cycles} & \colhead{km s$^{-1}$} & \colhead{km s$^{-1}$} & \colhead{} & \colhead{deg} & \colhead{HJD-2,400,000 d} & \colhead{$10^6$ km} & \colhead{$M_\odot$} & \colhead{km s$^{-1}$} & \colhead{}
}
\startdata
$ 0.441432 $ & $13334.0$ & $33.7$& $51.3$ & $0.166 $ & $258.42$& $56087.54$&  $0.307$& $0.00592$&$19.043$& $30$\\
$\pm 2e-06$ &  & $\pm 4.0$& $\pm 6.0$ & $\pm0.103$ & $\pm 36.38$& $\pm0.041$&  $\pm 0.036$& $\pm 0.00208$&&
\enddata
\tablecomments{Errors on each value are given in the second row.}
\end{deluxetable*}

\begin{deluxetable*}{cccccc}\label{table:4003_rv}
\tablehead{\colhead{WOCS ID} & \colhead{HJD - 2400000} & \colhead{Radial Velocity}& \colhead{O-C}& \colhead{Phase}& \colhead{Correlation Height}\\
\colhead{}& \colhead{days} & \colhead{$\rm{km\;s^{-1}}$}& \colhead{$\rm{km\;s^{-1}}$}& \colhead{}& \colhead{}}
\caption{WOCS radial velocity measurements of WOCS 4003 and 6038}
\startdata
   4003 & 53388.7347 &	82.0 &	-0.7 &	0.25 &	0.56\\
   4003 & 53720.8825 &	-13.5 &	-12.8 &	0.69 &	0.55\\
   4003 & 53807.6555 &	94.5 &	12.0 &	0.26 &	0.62\\
   4003 & 53811.8447 &	-13.0 &	-1.0 &	0.75 &	0.37\\
   4003 & 54100.9591 &	-14.1 &	-11.9 &	0.69 &	0.61\\
   4003 & 54164.7542 &	85.9 &	2.9 &	0.21 &	0.50\\
   4003 & 54165.7418 &	48.6 &	-4.9 &	0.45 &	0.63\\
   4003 & 54166.7401 &	-20.7 &	-15.1 &	0.71 &	0.54\\
   4003 & 54549.8257 &	28.7 &	-5.4 &	0.54 &	0.61\\
   4003 & 54602.7221 &	83.3 &	13.2 &	0.36 &	0.59\\
   4003 & 54604.6704 &	-14.0 &	2.0 &	0.78 &	0.61\\
   4003 & 55191.0099 &	46.6 &	5.5 &	0.04 &	0.66\\
   4003 & 55202.0440 &	48.8 &	9.4 &	0.04 &	0.66\\
   4003 & 56050.7182 &	28.8 &	6.7 &	0.59 &	0.62\\
   4003 & 56311.9980 &	62.2 &	14.6 &	0.48 &	0.53\\
   4003 & 56313.0099 &	-12.3 &	2.7 &	0.77 &	0.61\\
   4003 & 56350.9004 &	38.5 &	20.7 &	0.60 &	0.54\\
   4003 & 57034.0283 &	47.7 &	-24.1 &	0.13 &	0.66\\
   4003 & 57062.9167 &	39.7 &	14.4 &	0.57 &	0.65\\
   4003 & 57087.8819 &	50.7 &	-20.4 &	0.13 &	0.59\\
   4003 & 57405.9634 &	9.7 &	11.9 &	0.69 &	0.63\\
   4003 & 57407.0039 &	34.8 &	-9.7 &	0.05 &	0.65\\
   4003 & 57443.9111 &	29.2 &	23.9 &	0.66 &	0.63\\
   4003 & 57444.9016 &	-9.7 &	2.6 &	0.90 &	0.63\\
   4003 & 57762.8853 &	81.4 &	-1.5 &	0.25 &	0.63\\
   4003 & 58214.7878 &	2.8 &	-4.5 &	0.97 &	0.61\\
   4003 & 58563.1301 &	91.3 &	33.5 &	0.08 &	0.61\\
   4003 & 58565.0851 &	-22.0 &	-61.3 &	0.51 &	0.60\\
   4003 & 58857.0673 &	-3.3 &	-6.6 &	0.96 &	0.60\\
   4003 & 59274.8152 &	82.9 &	4.0 &	0.30 &	0.56\\
   6038 & 53720.8825 &	32.9 &	 &	 &	0.92\\
   6038 & 57087.8819 &	34.5 &	 &	 &	0.91\\
   6038 & 57407.0039 &	34.5 &	 &	 &	0.94\\
   6038 & 57448.8904 &	35.0 &	 &	 &	0.93\\
   6038 & 57762.8853 &	35.4 &	&	 &	0.93\\
    \enddata
\tablecomments{See \cite{gellerStellarRadialVelocities2021} for more information on data.}
    \label{tab:placeholder}
\end{deluxetable*}

\clearpage
\bibliography{references}{}
\bibliographystyle{aasjournalv7}

\end{document}